\documentclass[10pt,english,twocolumn]{revtex4-2}
\usepackage[T1]{fontenc}

\usepackage[utf8]{inputenc} 
\usepackage{geometry}
\usepackage{units}
\usepackage{textcomp}
\usepackage{amsmath}
\usepackage{babel}
\usepackage{graphicx}
\usepackage{xcolor}

\begin{document}


\author{M. Ran\v{c}i\'{c}$^{1a)}$, M. Le Dantec$^{1a)}$, S. Lin$^{2}$, S. Bertaina$^{3}$, T. Chaneli\`ere$^{4}$, D. Serrano$^{5}$, P. Goldner$^{5}$, R. B. Liu$^{2}$, E. Flurin$^{1}$, D. Est\`eve$^{1}$, D. Vion$^{1}$, P. Bertet$^{1}$}

\email{patrice.bertet@cea.fr}
\thanks{\\a) These authors contributed equally to this work.}


\affiliation{$^1$Universit\'e Paris-Saclay, CEA, CNRS, SPEC, 91191 Gif-sur-Yvette Cedex, France\\
$^2$Department of Physics, Centre for Quantum Coherence, and The Hong Kong Institute of Quantum Information Science and Technology, The Chinese University of Hong Kong, Shatin, New Territories, Hong Kong, China \\
$^3$CNRS,  Aix-Marseille  Universit\'e,  IM2NP  (UMR  7334),  Institut  Mat\'eriaux Micro\'electronique  et  Nanosciences de Provence,  Marseille,  France \\
$^4$Univ. Grenoble Alpes, CNRS, Grenoble INP, Institut N\'eel, 38000 Grenoble, France \\
$^5$Chimie ParisTech, PSL University, CNRS, Institut de Recherche de Chimie Paris, 75005 Paris, France \\
$^6$Walther Meissner Institut, Bayerische Akademie der Wissenschaften, Garching, Germany}

\title{Electron-spin spectral diffusion in an erbium doped crystal at millikelvin temperatures}

\begin{abstract}
Erbium-doped crystals offer a versatile platform for hybrid quantum devices because they combine magnetically-sensitive electron-spin transitions with telecom-wavelength optical transitions. At the high doping concentrations necessary for many quantum applications, however, strong magnetic interactions of the electron-spin bath lead to excess spectral diffusion and rapid decoherence. Here we lithographically fabricate a 4.4 GHz superconducting planar micro-resonator on a $\text{CaWO}_{4}$ crystal doped with Er ions at a concentration of twenty parts per million relative to Ca. Using the microwave resonator, we characterize the spectral diffusion processes that limit the electron-spin coherence of Er ions at millikelvin temperatures by applying 2- and 3-pulse echo sequences. The coherence time shows a strong temperature dependence, reaching 1.3 ms at 23 mK for an electron-spin transition of $^{167}\text{Er}$.
\end{abstract}

\maketitle

\section{Introduction}
Rare earth ion doped materials cooled to liquid helium temperatures
have demonstrated long coherence times \cite{Zhong2015a,Holzapfel2020a,Longdell2005,Rancic2017}
and large efficiencies for quantum memory demonstrations in the optical
domain \cite{Hedges2010,Dajczgewand2014,Sabooni2013}. Recently, interest
in these materials has extended to sub-Kelvin temperatures and sub-Tesla
fields, where the electron-spin transitions of Kramers (odd-electron)
rare-earth ions achieve gigahertz frequencies while exceeding the
energy of the thermal bath. This interest is motivated by the prospect of ensemble-based microwave-to-optical conversion \cite{Bartholomew2020,Fernandez2019} and microwave quantum memories \cite{Grezes2016,Sigillito2014}, both of which require long electron-spin coherence times and sufficient concentration for high-fidelity operation. Previous demonstrations of ensemble-based microwave quantum memories required concentrations of electron-spins between one and one hundred parts per million (ppm) to achieve storage efficiencies of order $10^{-3}$ \cite{Grezes2015, Ranjan2020, Probst2015}.

Several rare-earth doped materials have also demonstrated millisecond-scale electron-spin coherence \cite{Dold2020,Li2020} and hybridised electron-nuclear-spin coherence \cite{Rakonjac2020,Ortu2018} with these applications in mind. Amongst these proposed materials, erbium doped calcium tungstate ($\text{Er}^{3+}\text{:CaWO}_{4}$) has emerged as a leading candidate due to its telecom-wavelength optical transition and $\sim$20 ms coherence both predicted \cite{Dantec2021, Kanai2021} and measured \cite{Dantec2021} on a magnetically-sensitive electron-spin transition. This demonstration of long coherence was attributed to a millikelvin spin-bath temperature combined with an ultra-low Er doping concentration of just 0.7 ppb, yielding a regime in which the weak magnetic interaction with the $^{183}$W nuclear-spin bath dictated the electron-spin decoherence rate.

At higher doping concentrations that are required for quantum applications, however, the strong magnetic interactions between the electron-spins of erbium ions are expected to induce faster decoherence due to spectral diffusion (SD) \cite{Herzog1956}. Here we use pulsed electron spin resonance (ESR) spectroscopy to investigate SD in 20 ppm Er doped $\text{CaWO}_{4}$ in the millikelvin temperature regime. Moreover, we demonstrate a two-pulse echo coherence time of 1.3 ms for an electron-spin transition of Er at 4.4 GHz and determine that this coherence is limited by unwanted paramagnetic impurities, indicating that $\text{Er}\text{:CaWO}_{4}$ could be suitable for quantum information processing applications requiring large optical or microwave absorption.

\section{Structure of $\text{Er}^{3+}\text{:CaWO}_{4}$}
Calcium tungstate is an optically-transparent crystal with tetragonal unit cell structure (space-group $I{4}_{1}/a$) and orthogonal crystal axes $a,b,c$. An illustration of the unit cell structure is presented in Figure 1a, which has dimensions $5.2 \times 5.2 \times 11.4$ $\text{\AA}$ parallel the crystal $a \times b \times c$ axes. Crystals grown with a natural abundance of isotopes exhibit an intrinsically low level of magnetic noise, originating almost exclusively from the nuclear spins of the $^{183}$W isotope. This isotope has a natural abundance of 14 $\%$ and a nuclear spin-half moment with a relatively small gyromagnetic ratio of $\gamma_{W}/2 \pi = 1.8$ MHz/T. For this reason $\text{CaWO}_{4}$ is considered one of the best candidate hosts for quantum information processing amongst thousands of known materials \cite{Kanai2021, Ferrenti2020}. 

In the crystalline matrix, the $\mathrm{Er}^{3+}$ ions substitutionally replace $\mathrm{Ca}^{2+}$ ions with additional long-range charge compensation  \cite{Kiel1970}. The electrostatic interaction between the Er ions and the $\text{CaWO}_{4}$ matrix lifts the 16-fold degeneracy of the $J=15/2$ electronic ground state of trivalent erbium, leading to eight pairs of degenerate electronic sub-levels known as Kramers doublets. Each doublet forms an effective spin-1/2 system and only the lowest energy Kramers doublet is populated at millikelvin temperatures. In the presence of an applied magnetic field $B_0$, this effective electron-spin shows a strongly anisotropic Zeeman effect, characterized by the Hamiltonian:
\begin{alignat*}{1} 
H_Z = \mu_B\,\mathbf{S\cdot g \cdot B_0.}  
\end{alignat*} Here $\mu_{B}$ is the Bohr-magneton and $\mathbf{g}$ is the g-tensor whose symmetry mirrors the tetragonal $S_4$ point-group symmetry of the crystal electric-field at the location of the Er ion \cite{Bertaina2007}.
\begin{alignat*}{1}
\mathbf{g} & =\left[\begin{array}{ccc}
8.38 & 0 & 0\\
0 & 8.38 & 0\\
0 & 0 & 1.247
\end{array}\right]_{\left(a,b,c\right)}
\end{alignat*}
Additionally, 23 \% of Er ions belong to the $\text{}^{167}\text{Er}$ isotope which has non-zero nuclear-spin $\mathbf{I}=7/2$. For this subset of ions the magnetic-hyperfine interaction is parameterised by the hyperfine $\mathbf{A}$-tensor and therefore an additional term is required in the effective-spin Hamiltonian:
\begin{alignat*}{1}
H_{\text{Er-167}} & =\mu_{B}\,\mathbf{S\cdot g\cdot B_{0}}+\mathbf{S\cdot A\cdot I.}
\end{alignat*}
Note that we have neglected the weak magnetic-nuclear and nuclear-quadrupole interactions in the $\text{}^{167}\text{Er}$ spin-Hamiltonian \cite{Guillot-Noel2006b} because the additional precision awarded by these terms is not required to describe the processes studied here.
Meanwhile, the $\mathbf{A}$-tensor once again reflects the symmetry of the $\text{CaWO}_{4}$ matrix with $A_{aa}=A_{bb}=-873$ MHz, $A_{cc}=-130$ MHz and the Hamiltonian ${H}_{\text{Er-167}}$ yields a total of 16 energy levels. The energy level diagrams of both $H_Z$ and ${H}_{\text{Er-167}}$ are presented in the lower panel of Figure 2 for small magnetic fields applied perpendicular to the crystal $c$-axis.

\begin{figure}[tbh!]
\centering
\includegraphics[width = 8cm]{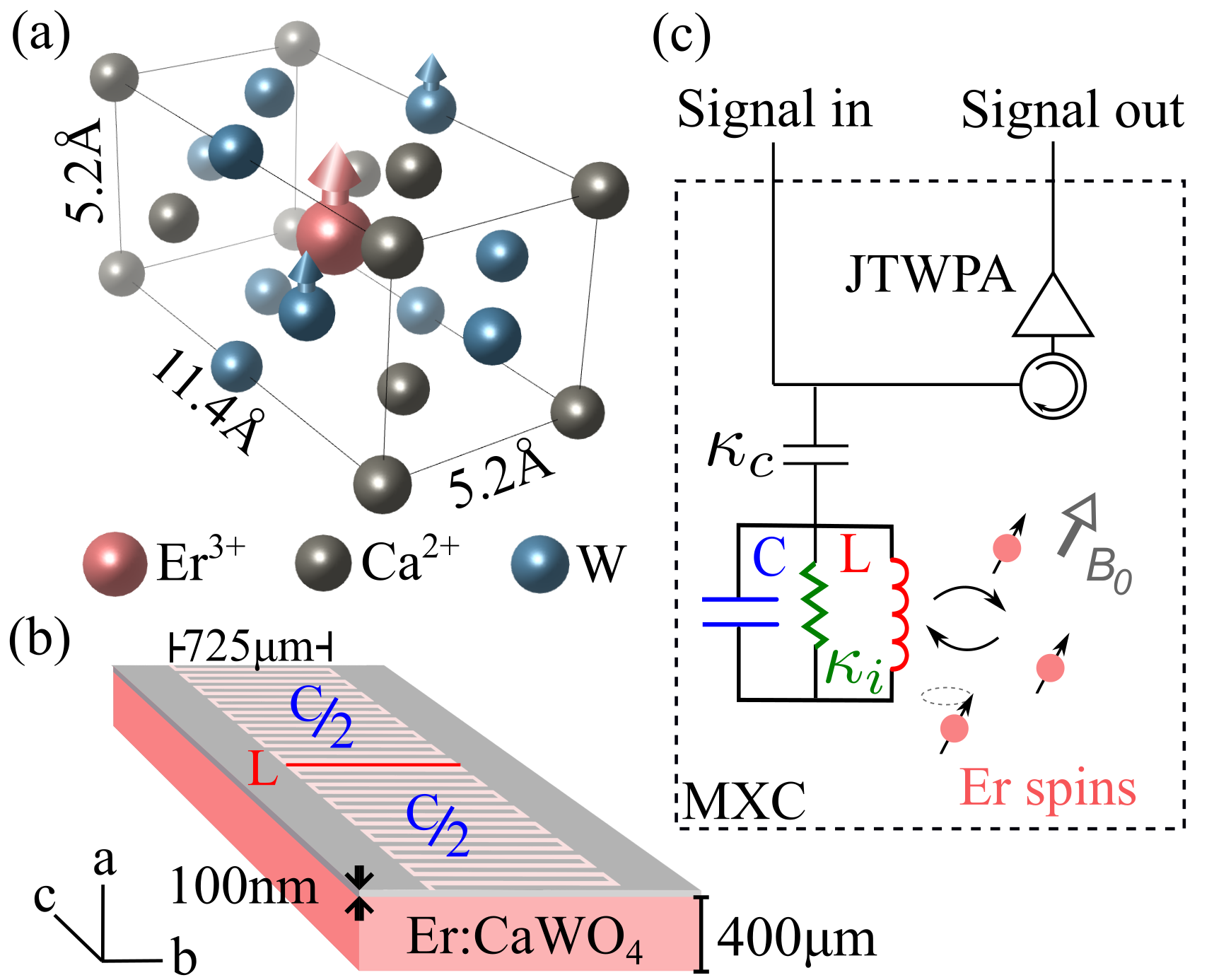}
\caption{Experimental setup. (a) Unit cell of $\textrm{CaWO}_{4}$ showing a central Er dopant. Oxygen atoms removed for clarity (b) 3D perspective view of $\mathrm{Er}^{3+}:\mathrm{CaWO}_{4}$ crystal, showing the thin-film Nb resonator patterned on the surface. Crystal axes relevant to sub-figures (a) and (b) are shown in bottom left corner. (c) Experimental ESR setup, comprising of the resonator and JTWPA. }
\label{fig1}
\end{figure}

\section{Experimental Setup}

For these measurements we use an ESR spectrometer
comprised of a thin-film superconducting resonator and Josephson Traveling Wave Parametric Amplifier (JTWPA) \cite{Macklin2015}. Such spectrometers have been described in detail in refs \cite{Probst2017,Bienfait2015a} and achieve sensitivities as high as $12\,\mathrm{spins}/\sqrt{\mathrm{Hz}}$ for detecting two-pulse (Hahn) echoes originating from donors in silicon at mK temperatures \cite{Ranjan2020}. This high sensitivity is due to the very low output noise, which is dominated by quantum fluctuations of the microwave field with little contribution from thermal photons at millikelvin temperatures. A 3D perspective representation of the sample studied here is shown in Fig. 1 (b). The circuit is etched into a 100 nm Nb layer sputtered directly on the surface of the $\textrm{CaWO}_{4}$ crystal, which was grown by Scientific Materials Corp. and has dimensions $0.4\times3\times6$ mm parallel to the $a\times b\times c$ axes of the crystal. The superconducting LC resonator comprises 15 interdigitated fingers (the capacitor) on either side of the 725 $\mu$m $\times$ 5 $\mu$m wire (the inductor). 

The frequency of the fundamental resonator mode is $\omega_{r}/2\pi$ = 4.37 GHz in the absence of a magnetic field, with a quality factor $Q=8\cdot10^{3}$. The resonator $Q$ is determined by the coupling rate to the measurement line $\kappa_{C}=3\cdot10^{6}\,s^{-1}$ and the internal energy loss rate $\kappa_{i}=5\cdot10^{5}\,s^{-1}$. For these experiments a DC magnetic field $B_{0}$ is applied parallel to the sample surface in the direction of the inductance wire (the crystal $b$-axis). Precision alignment of the magnetic field with respect to the crystal surface is important for minimising the internal loss rate $\kappa_{i}$ because field penetration into the superconducting thin-film generates magnetic vortices; a well-known microwave loss channel \cite{Song2009}.

To achieve this precision alignment, the sample is enclosed within a small copper box and mounted onto two copper-beryllium actuators from Attocube, comprising a goniometer and rotator. These are inserted into a set of two Helmholtz coils, and thermally anchored to the mixing chamber (MXC) of a dilution refrigerator. While the actuators precisely orient the sample with respect to the applied magnetic field, they also create a weak thermal bridge due to the separation of metallic components by their piezo-electric elements and therefore additional copper-braiding is used between the sample-box and MXC assembly to maintain good thermal conductivity. 

Erbium electron-spin transitions are excited with microwave signals sent from a heavily attenuated (~50 dB) input line. The transmitted microwave signal, together
with the signal emitted by the spin ensemble, is then amplified by the JTWPA (see Fig. 1b). Further microwave amplification takes place at 4K using a high electron mobility transistor (HEMT) and at room temperature using low-noise semiconductor amplifiers.

\section{ESR spectroscopy}

\begin{figure*}[tbh!]
\centering
\includegraphics[width = 13cm]{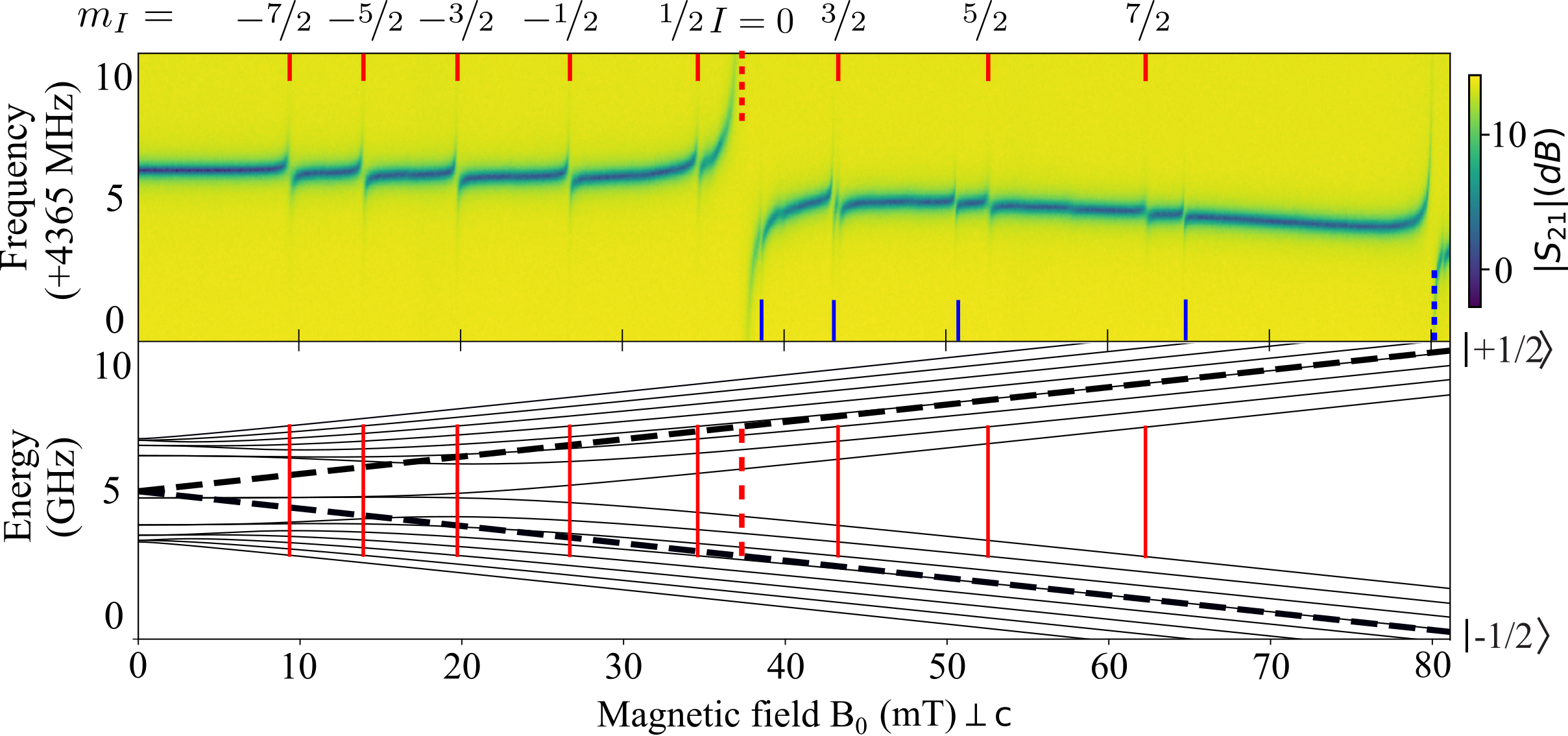}
\caption{\textit{Top:} ESR spectrum measured at 100 mK. At this temperature all 8 ESR transitions of $^{167}\text{Er}$ are populated, and these are labelled and indicated with vertical red lines at the top of the spectrum. The $m_I=+3/2$ transition studied here is just right of the large Er $I=0$ transition, marked by vertical red dashes. Five additional ESR transitions of Yb are indicated with vertical blue lines at the bottom of the spectrum. The Yb $I=0$ transition at 80 mT (blue dashed line) is particularly important due to its SD contribution at low temperatures. \textit{Bottom:} Energy level diagram of  $^{167}\text{Er}$ (solid black lines) and of Er $I=0$ (black dashed lines). The energy levels of $^{167}\text{Er}$ are designated by spin-state projections $|m_S, m_I\rangle$ with the two $m_S$ multiplets labelled on the right. The ESR transitions of Er identified in the spectrum above are marked correspondingly in red. \label{fig2}}
\end{figure*}

This crystal demonstrates a rich ESR spectrum due to the existence of several Er isotopes in natural abundance. Shown in Figure 2, the microwave transmission spectrum is measured as a function of magnetic field applied along the crystal $b$-axis. Note that this spectrum was recorded at an elevated temperature of 100 mK in order to populate all the hyperfine levels of ${}^{167}\mathrm{Er}$.

As stated previously, however, most isotopes of Er have no nuclear spin and the largest anti-crossing at 37 mT is associated with the high concentration of $I=0$ erbium isotopes; predominantly ${}^{166}\mathrm{Er}$, ${}^{168}\mathrm{Er}$ and ${}^{170}\mathrm{Er}$. Here the cooperativity $C$ between the resonator and the spin-ensemble is large $\left(C\approx50\right)$ due to the high spin density. Unfortunately, this high density of spins also gives rise to large instantaneous diffusion (ID), making it difficult to measure SD using the $I=0$ transition (see Sup. Mat. Section I). To avoid this technical limitation, we instead focus on the low-density ${}^{167}\mathrm{Er}$ isotope, which exhibits 16 hyperfine levels. The ESR-allowed $\Delta m_{I}=0$ transitions between these levels are labelled in Fig. 2 according to their nuclear-spin projection $|{m}_{I}\rangle$. The 8 corresponding solid red lines indicate the values of $B_{0}$ for which the ESR frequencies of ${}^{167}\mathrm{Er}$  are equal to $\omega_{r}$, both experimentally (Fig. 2 Top) and theoretically (Fig. 2 Bottom). Avoided level crossings are observed in the spectrum at these field values, which is consistent with high-cooperativity between the resonator and each transition \cite{Probst2013a,Kubo2010}. Additionally, the five avoided level crossings observed in the spectrum above 38 mT are attributed to Yb impurities. The large anti-crossing at 80 mT is attributed to the $I=0$ Yb isotopes, comprising of all the even-numbered isotopes between ${}^{168}\mathrm{Yb}$ and ${}^{176}\mathrm{Yb}$.


\section {\label{sec:SD}Spectral Diffusion Model}

We use both two-pulse echoes (2PE) and three-pulse echoes (3PE) to
elucidate the dynamic interactions which cause SD within the $m_{I}=|+\nicefrac{3}{2}\rangle$ spin-ensemble of the ${}^{167}\mathrm{Er}$ isotope. Here the 2PE consists of a $\nicefrac{\pi}{2}$-pulse followed by a $\pi$-pulse at time delay $\tau$, while the 3PE consist of three $\pi/2$ pulses with separations $\tau$ and $T_{W}$ between pulses 1-2 and 2-3, respectively. The 2PE and 3PE data yield complementary information, and were first used in the 1960's by Mims, Nasau and McGee to study SD of paramagnetic rare-earth-ion impurities \cite{Mims1961}. 

Since then, the theory of spectral diffusion has evolved considerably, and here we fit the 2PE and 3PE data using the uncorrelated-sudden-jump model first described by Hu and Hartmann in 1974 \cite{Hu1974}. This model is known to yield accurate fits of optical 2PE and 3PE data in rare-earth-ion doped materials \cite{Bottger2006} and relies on four key assumptions:

1) The subset of the $m_{I}=|+\nicefrac{3}{2}\rangle$ transition under investigation (the excited spins) are sufficiently isolated from each other such that they do not contribute to SD. 

2) SD is instead caused exclusively by the magnetic dipole interaction with other denser baths of perturbing spins. 

3) These perturbing spins are randomly located within the crystal matrix and exhibit only two energy configurations. i.e: a Spin-1/2 system

4) Spin-flips within the baths of perturbing spins are uncorrelated and so each flip is treated independently.
To describe explicitly this model of SD we utilise the formalism first introduced by Bai and Fayer \cite{Bai1989} and later summarised by Bottger et. al. in a single equation suitable for both 2P- and 3P-echoes \cite{Bottger2006}. From this model, we derive a similar echo amplitude decay $A_{E}$ as a function of the inter-pulse delays, extended to account for several species S of perturbating spins:
\begin{alignat}{1}
\text{A}_{E} & =\text{V}_{eseem}\cdot\text{A}_{0}\exp\left(-2\pi\Gamma_{eff}\tau\right)\exp\left(\frac{-T_W}{T_1}\right)\label{eq: Generic Echo}\\
\Gamma_{eff} & =\Gamma_{0}\underset{S\quad\,}{+\sum\frac{\Gamma_{\text{SD}}^{S}}{2}}\left\{ R^{S}\tau+1-\exp\left(-R^{S}T_{W}\right)\right\}. \label{eq:Gamma_effective}
\end{alignat} 

Here $A_0$ represents the limiting echo amplitude at zero delay and $\text{V}_{eseem}(\tau,T_{W})$ corresponds to the Electron Spin Echo Envelope
Modulation (ESEEM) caused by the magnetic dipole-dipole interaction of
the Er electron-spins and the proximal $^{183}\text{W}$ nuclear-spins. This
is a well defined temporal modulation that depends only on the orientation
and magnitude of the applied magnetic field, and has already been
determined experimentally for the $m_{I}=|+\nicefrac{3}{2}\rangle$ electron-spin transition studied here \cite{Probst2020a}. 

Meanwhile, the decaying component of the echo signal contains the coefficients $\Gamma_{0}$, $\Gamma_{\text{SD}}^{S}$, $R^S$ and $T_1$. The coefficient $\Gamma_{0}$ represents decoherence processes
which occur on faster timescales than the measurement and instantaneous
diffusion (ID) caused by the microwave excitation pulses. This is equivalent to the homogeneous linewidth of a single spin transition in the absence of spectral diffusion. Conversely, $\Gamma_{\text{SD}}^{S}$ and $R^S$ describe the spectral diffusion which occurs during the delays $\tau$ and $T_W$ due to spin species $S$. More specifically, $R^S$ represents the average spin-flip rate and $\Gamma_{\text{SD}}^{S}$ is the full-width-half-maximum (FWHM) contribution to the dynamic distribution of transition frequencies within the $m_{I}=|+\nicefrac{3}{2}\rangle$ sub-ensemble.

Indeed, any paramagnetic species or ESR transition that exhibits ppm or higher concentration can contribute significant amounts of spectral diffusion, and therefore it is important to identify all paramagnetic impurities in the material and determine their concentration. Initially, this was attempted via magnetic-rotation-spectroscopy (Sup. Mat. Section II) combined with numerical spin-resonator coupling estimates (Sup. Mat. Section III) which identified both $\text{Yb}^{3+}$ and $\text{Mn}^{2+}$ impurities. 

However, our quantum-limited spectrometer was not able to detect paramagnetic impurities with small g-factors or long spin-lattice relaxation rates due to the limited magnetic field (400 mT) and temperature (600 mK). Thus a piece of the $\text{CaWO}_{4}$ boule was submitted for inductively-coupled-plasma mass-spectrometry (ICP-MS) analysis and the results are presented in Table \ref{tab:mass-spec} \cite{An}. 

Mass-spectrometry allowed for the concentration of the previously-identified Er, Yb and Mn impurities to be determined with high precision and three additional rare-earth-ion impurities were also identified in the crystal at lower concentrations.

\begin{table}[ht]
\begin{centering}
\begin{tabular}{|c|c|c|c|c|c|c|}
\hline 
Element & Mn & Ce & Nd & Gd & Er & Yb\tabularnewline
\hline 
\hline 
Conc. $\left(\mu g/g\right)$ & 0.542 & 0.424 & 0.211 & 0.066 & 13.292 & 8.516\tabularnewline
\hline 
Conc. (ppm) & 2.840 & 0.871 & 0.421 & 0.120 & 22.88 & 14.17\tabularnewline
\hline 
\end{tabular}
\par\end{centering}
\caption{\label{tab:mass-spec}Calcium-relative impurity concentrations detected in a dissolved (aqueous) sample of the $\text{CaWO}_{4}$ boule, represented by relative mass concentration (top row) and the equivalent relative doping concentration (bottom row). This measurement was performed at the European Centre for Environmental Geoscience Research and Teaching using a Perkin Elmer NexION 300X mass-spectrometer.}
\end{table}

The two ESR transitions with the greatest absorption; the $I=0$ Er and Yb transitions, were thereby included in the SD analysis. For each transition (i.e: species) $S$, the linewidth $\Gamma_{\text{SD}}^{S}$ is proportional to the number of spins which can undergo spin-flip, and the expected temperature dependence is described by Boltzmann statistics \cite{Bottger2006}:
\begin{alignat}{1}
\Gamma_{\text{SD}}^{S}(T) & =\Gamma_{\text{max}}^{S}\text{sech}^{2}\left(\frac{g_{S}\mu_{B}B_{0}}{2kT_B}\right),\label{eq:Gamma_SD}
\end{alignat}
where $k$ is the Boltzmann constant, $T_B$ is the electron-spin bath temperature and
$g_{S}$ is the g-factor of the electron-spin transition in the direction of the applied field $B_{0}$. The constant $\Gamma_{\text{max}}^{S}$
is the maximum dynamic linewidth contribution from species $S$.
In the high-temperature limit $\Gamma_{\text{SD}}^{S}$ approaches $\Gamma_{\text{max}}^{S}$ asymptotically, as the spin-down and spin-up populations of species $S$ equalize and the magnetic-dipole interaction with the $m_{I}=|+\nicefrac{3}{2}\rangle$ sub-ensemble is maximised. Meanwhile,
$R^{S}$ is the associated rate of spin-flips (the sum of upward and
and downward spin-flips) which is also temperature dependent and has two contributions for the studied
temperature range. The first is due to the spin flip-flops, which follows
the same temperature dependence as in Eq. \ref{eq:Gamma_SD}. The
second is the spin-lattice (phonon) interaction resonant with the
spin-transition frequency $\omega_{S}=\mu_{B}g_{S}B_{0}/\hbar$, yielding the following equation \cite{Bottger2006, Dantec2022}:
\begin{alignat}{1}
R^{S}\left(T\right) & =\alpha_{ff}\frac{{g_{S\perp}}^{4}n_{S}^{2}}{\Gamma_{S}}\text{sech}^{2}\left(\frac{g_{S}\mu_{B}B_{0}}{2kT_B}\right)\qquad\label{eq:R}\\
 & \qquad+\alpha_{ph}g_{S}^{3}B_{0}^{5}\coth\left(\frac{g_{S}\mu_{B}B_{0}}{2kT_B}\right).\nonumber 
\end{alignat}Here $\alpha_{ff}$ and $\alpha_{ph}$ represent scaling
constants of the flip-flop and spin-lattice interactions respectively, and it is assumed they do not vary between different paramagnetic species in the crystal. Meanwhile, $n_{S}$ represents the density of species $S$ and $\Gamma_{S}$ the FWHM of the inhomogeneously
broadened ESR absorption line, which can depend on both the magnitude and direction of the applied magnetic field. Additionally, $g_{S}$ represents the g-factor in the direction of the applied field, while $g_{S\perp}$ is the g-factor in the direction orthogonal to the c-axis of the crystal. Note that the flip-flop rate is only sensitive to the components of the g-tensor perpendicular to the direction of the applied field \cite{Car2019}. Here the magnetic field is applied along the $b$-axis of the crystal, so this coupling depends on the spectroscopic splitting factors in the directions of both the a- and c-axes, namely $g_{\perp}$ and $g_{\parallel}$. As $g_{\parallel} \ll g_{\perp}$ for both erbium and ytterbium, we approximate this angular dependence using only $g_{\perp}$ in the expression for the flip-flop rate.

In the limit $T_{W}=0$, the effective linewidth presented in Eq (\ref{eq:Gamma_effective}) simplifies to $\Gamma_{0}+\nicefrac{\tau}{2}\sum\Gamma_{\text{SD}}^{S}R^{S}$. Thus, it is impossible to obtain independent values for $\Gamma_{\mathrm{SD}}$ and $R$ with only 2PE measurements, due to the multiplicative relationship between these two parameters.
For this reason 3PE measurements are necessary to obtain an independent fit of $\Gamma_{\mathrm{SD}}$ and $R$.
\begin{figure}[tbh]
\centering
\includegraphics[width = 8cm]{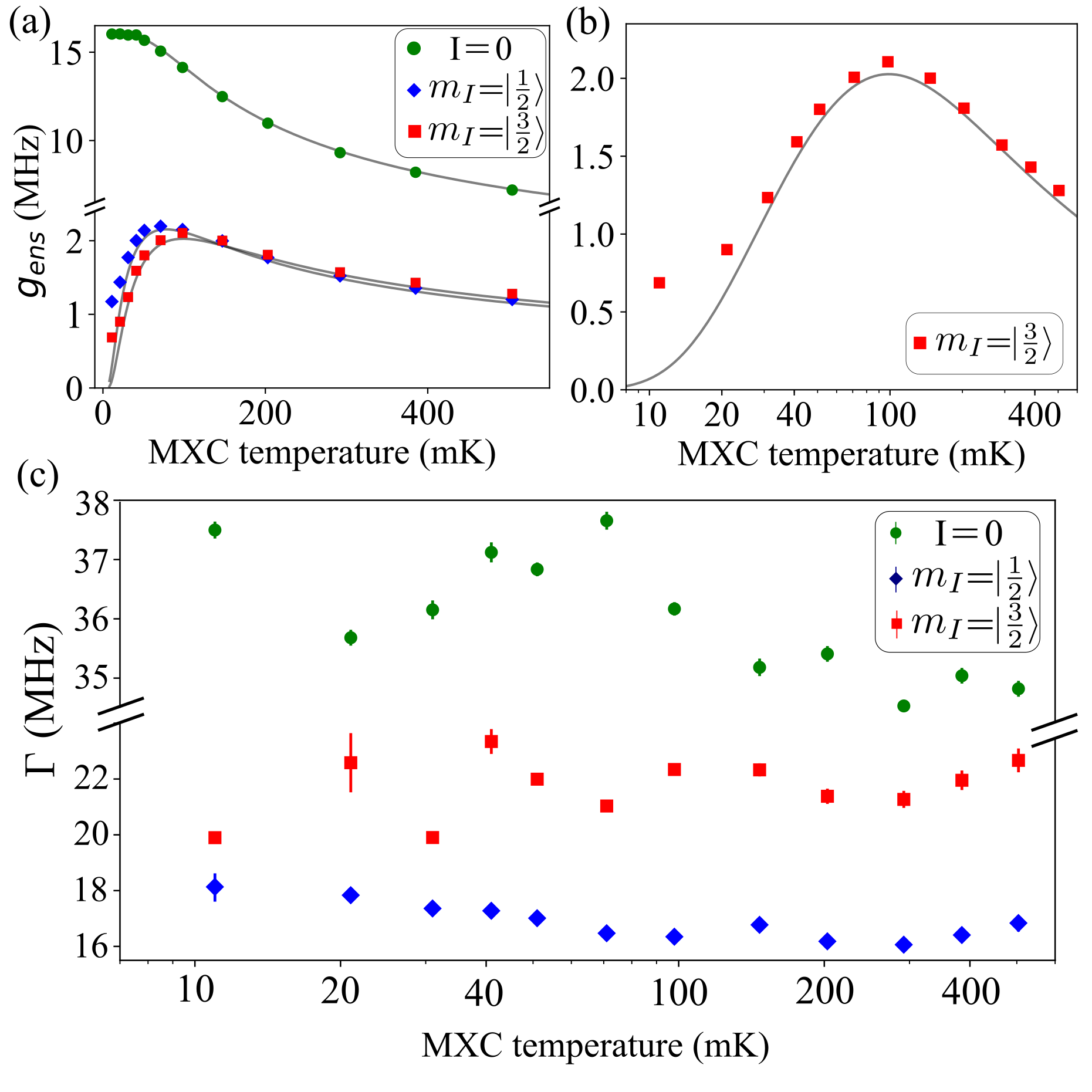}
\caption{Ensemble coupling and inhomogeneous linewidths as a function of MXC temperature, determined using a fit to Eq. (\ref{eq:S21}) for three electron-spin transitions of Er: The $I=0$ transition (green), the  $m_I=|\frac{1}{2}\rangle$ (blue) and $m_I=|\frac{3}{2}\rangle$ (red) transitions of $^{167}\mathrm{Er}$. (a) $g_{ens}$ values extracted from the fit (coloured dots). The solid black lines show a simultaneous fit to all three $g_{ens}$ data-sets assuming a thermal Boltzmann distribution of spin-levels. (b) Same data as in subplot (a) for the $m_I=|\frac{3}{2}\rangle$ transition, presented with a logarithmic temperature scale. (c) Inhomogeneous linewidths $\Gamma$ extracted from the same fitting as in subfigure (a) presented with a logarithmic temperature scale. Here a negligible temperature dependence is observed for all three transitions, consistent with the observation that inhomogeneous broadening of paramagnetic impurities is dominated by electrostatic perturbations in $\text{CaWO}_{4}$ \cite{Mims1966,Baibekov2014}.} 
\label{fig3}
\end{figure}

\section {\label{sec:Ens_coupling}Ensemble Coupling measurements}

To accurately fit the temperature dependence of $\Gamma_{\mathrm{SD}}$
and $R$ it is also important to know the temperature $T_B$ of the
electron-spin bath, as this can be higher than the temperature of the
mixing chamber $\left(T_{MXC}\right)$ due to limited thermal conduction. Here
we independently determine $T_B$ as a function of $T_{MXC}$ by evaluating the coupling of the $m_{I}=|+\nicefrac{3}{2}\rangle$ sub-ensemble to the resonator at each measured temperature, using a well established complex transmission formula \cite{Ranjan2013}:
\begin{alignat}{1}
S_{21}(\omega)= & 1-\frac{\kappa_{C}}{2i\left(\omega-\omega_{r}\right)+\kappa_{T}+\underset{S}{\sum}\frac{2{(2\pi g_{ens}^{S})}^{2}}{i\left(\omega-\omega_{S}\right)+\pi\Gamma_{S}}},\label{eq:S21}
\end{alignat}
where $g_{ens}^{S}$ is the temperature-dependent ensemble coupling between the electron-spin species $S$ and the resonator and the other variables are as previously defined. Example fits of $g_{ens}^S$ and $\Gamma_S$ are presented in section III of the supplementary materials for several Er and Yb transitions. Figure 3 presents the results of this fitting as a function of $T_{MXC}$ for three of the electron-spin transitions presented in Fig. 2; the $|+\nicefrac{1}{2}\rangle$, $I=0$ and $|+\nicefrac{3}{2}\rangle$ transitions.

Noting that $g_{ens}^{S}$ is proportional to the square-root of the electron-spin polarisation of species $S$ \cite{Ranjan2013}, we can directly compare the fitted ensemble-coupling at each measured temperature with the expected temperature dependence, assuming the Er electron- and nuclear-spin populations thermalise to $T_{MXC}$ according to Boltzmann statistics (refer to Sup. Mat. Section IV for more detail).

Indeed, the solid black lines in Fig. 3a represent a simultaneous fit to $g_{ens}^{S}$ for all three transitions. The only free parameter in this fit is the total Er concentration, which yields $2\cdot{10}^{17}\,\text{spins}/\text{cm}^3$, equivalent to 18 ppm substitutional doping. This is consistent with the concentration determined by mass-spectrometry (Table \ref{tab:mass-spec}) which was used for the analysis of the 3PE measurements (Sup. Mat. Section V). 

Figure 3b presents an enlarged view of the ensemble-coupling data and fit for the $m_I=|+\nicefrac{3}{2}\rangle$ transition. This transition was chosen to probe the spin-temperature below 100 mK because the nuclear-spin distribution of $^{167}\text{Er}$ is considerably more sensitive to heating in this temperature region than the electron-spin distribution of the $I=0$ Er isotopes, shown by the green curve in Fig. 3a. Moreover, we assume that this measurement of the hyperfine level temperature of $^{167}\text{Er}$ accurately determines the electron-spin bath temperature $(T_{B})$ of all paramagnetic impurities in the sample because of the strong hyperfine coupling between the electronic and nuclear spin of $^{167}\text{Er}$, which keeps both spin baths isothermal.
In particular, the fit to the data in Fig. 3b suggests that $T_{B}$ is well approximated by $T_{MXC}$ above 30 mK. Below this MXC temperature, however, we infer that poor thermal conduction combined with some heating at-or-near the location of the sample leads to an elevated bath temperature with $T_B=23\pm2$ mK at $T_{MXC}$ = 11 mK and $T_B=27\pm2$ mK when $T_{MXC}$ = 21 mK.

Both the excess heating and poor conductivity are likely caused by the piezo-electric actuators that precisely align the sample in the applied magnetic field. The issue of poor thermal conductivity was mentioned previously in Section II and the frictional `stick-slip' mechanism which moves the actuators could potentially be a heat-source even during static (non-moving) operation \cite{Zech2019}. This is possible if vibrations from the pulse-tube cooler couple to the loose frictional surfaces within the actuators, which are required for precise displacement. Indeed, thermal excursions up to 100 mK were observed when the actuators are actively displaced using a voltage signal, and this temperature rise is attributed to frictional heating.

\begin{figure}[ht]
\centering
\includegraphics[width = 8cm]{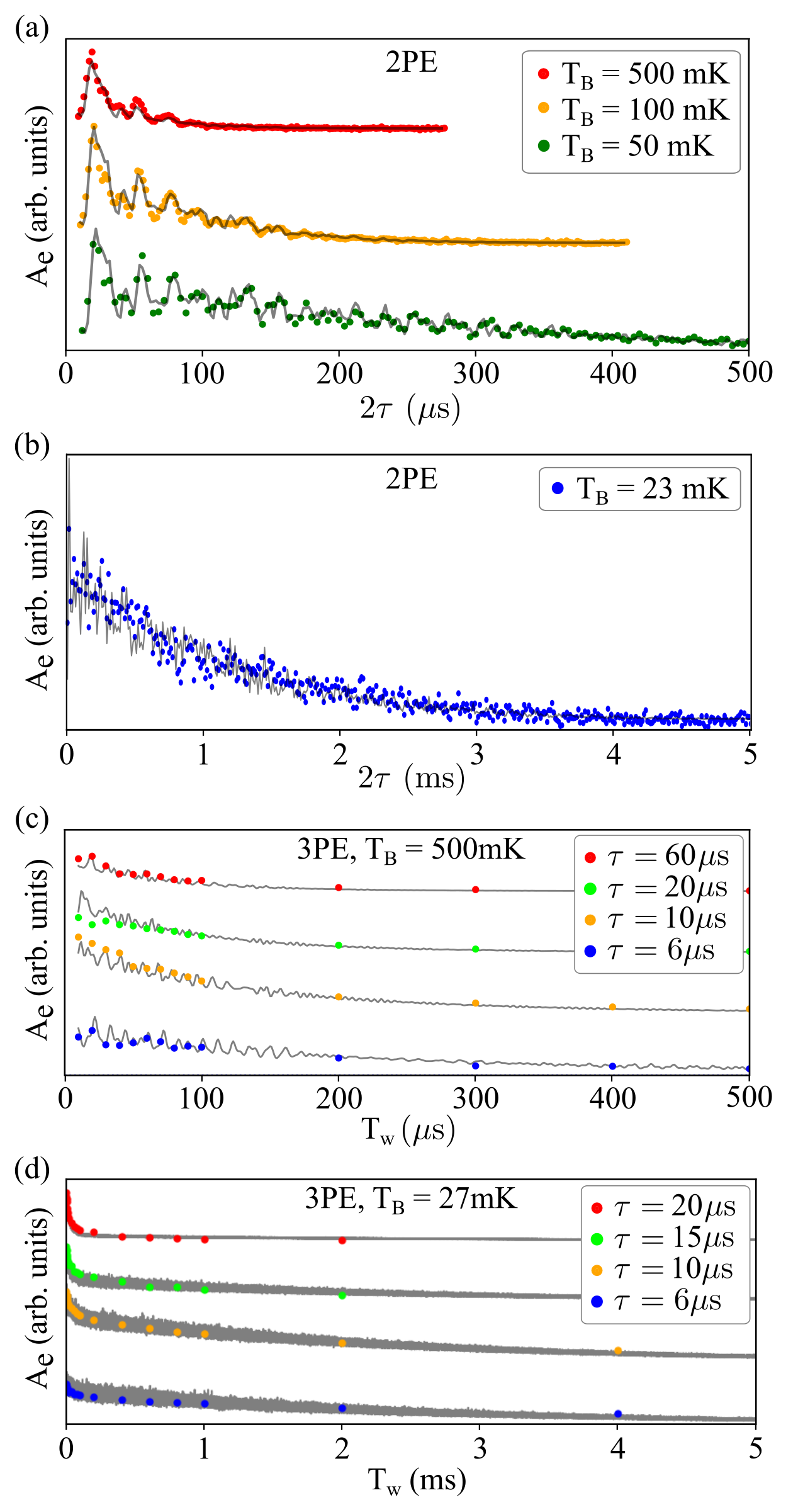}
\caption{Two- and three-pulse echo decay curves (coloured dots) measured as a function of delay times $\tau$ or $T_W$ and their respective fits (grey solid curves). (a) Waterfall plot of two-pulse echo decay-curves measured in the temperature range of 50-500 mK. (b) Two-pulse echo measurement recorded at $T_B = 23 $ mK; the lowest achievable spin-bath temperature for the experiments performed here. (c) Waterfall plot of three-pulse echo decay curves measured at $T_{B}=530$ mK as a function of both $\tau$ and $T_W$ time-delays. (d) Waterfall plot of three-pulse echo decay curves measured at $T_{B}=27$ mK.}
\label{fig4}
\end{figure}

\section{\label{sec:Echoes}Echo measurements}

As mentioned at the beginning of Section \ref{sec:SD}, 2PE and 3PE measurements were recorded at various spin-bath temperatures $T_B$ to determine the SD parameters $\Gamma_{\text{SD}}^{S}$ and $R^S$. A selection of these measurements is presented in Figure 4. To accurately determine the echo decay rates, it was necessary to fit the large ESEEM modulation depths observed in Fig. 4a and b. Fortunately, the ESEEM parameters have already been measured and published for the sample and experimental configuration used here \cite{Probst2020a}.

For the 3PE measurements, up-to ten decay curves were taken at five different temperatures as a function of  $T_{W}$, with fixed $\tau$  ranging from 6 $\mu s$ to 96 $\mu s$. These small values of $\tau$ were chosen for the 3PE measurements because they allow for accurate estimates of the rapid SD at higher temperatures. An example set of 3PE curves recorded at 530 mK and 27 mK are presented in sub-figures 4c and 4d, respectively. ESEEM is also present and simulated for the 3PE decay-curves, however, the data was recorded with large $T_{W}$-spacings and so the fitted curves in grey show an under-sampling of the modulation. 


\begin{figure}[ht]
\centering
\includegraphics[width = 8cm]{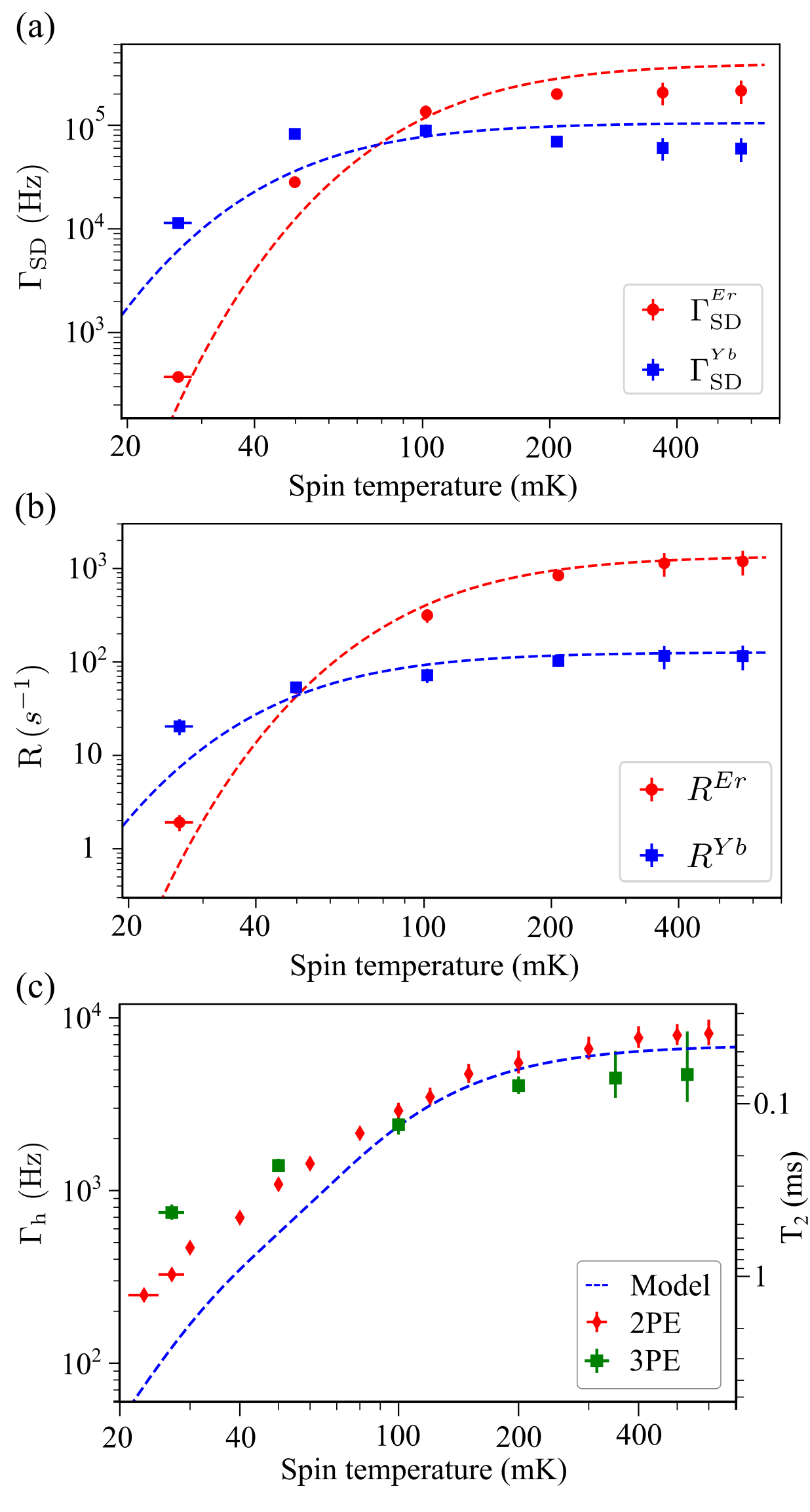}
\caption{Spectral diffusion and decoherence rates. (a) Linewidth $\Gamma_{\mathrm{SD}}$ and (b) spin-flip rates $R^{S}$ extracted from 3PE fits to Eq.(\ref{eq: Generic Echo}) for six temperatures and two spin species. Y-errors shows standard error in the fit. Red(blue) dashed lines in (a) and (b) represent fits to Eq. (\ref{eq:Gamma_SD}) and (\ref{eq:R}) for Er(Yb), respectively. (c) Extracted coherence time $T_2$ (Right hand side) and equivalent effective decoherence rate $\Gamma_h = 1/\pi{T}_{2}$ (left hand side) from the 2PE data (red) and the 3PE data (green). The decoherence model (in blue) takes into account the SD parameters extracted from the 3PE measurements, the ID of the $m_{I}=|+\nicefrac{3}{2}\rangle$ transition and the dephasing caused by the unpolarised ${}^{183}$W nuclear spin bath.}
\label{fig5}
\end{figure}

As mentioned previously, we rely on 3PE data to determine $R$ and $\Gamma_{\mathrm{SD}}$ at each measured temperature because 2PE measurements cannot be used to fit these these two parameters independently. For this fitting we also consider only the flip-flop contribution of $R^{S}\left(T\right)$ and the relative concentration of Er and Yb impurities. The result of this fitting is presented in Figures 5a and 5b, and the details of the fitting process are presented in section V of the supplementary materials.
For both the 2PE and 3PE measurements it is then possible to define a coherence time 
\begin{alignat}{1}
T_{2} & = \frac{-a+\sqrt{a^{2}+2b/\pi}}{b}.
\end{alignat}
Here $a=\Gamma_{0}$ and $b=\frac{1}{2}\sum_{S}\Gamma_{\mathrm{SD}}^{S}R^{S}$.
These two parameters arise from the quadratic exponent $\Gamma_{eff}\tau=a\tau+b\tau^{2}$ presented in Eq. (\ref{eq:Gamma_effective}) when $T_{W}=0$. For both the 2PE and 3PE measurements, this value of $T_2$ represents the time $2\tau$ when the echo has decayed to $e^{-1}$ of its initial amplitude (in the limit $T_{W}=0$ for 3PE measurements) and is presented in Fig. 5c. 

Above 100 mK the value of $T_2$ inferred from either 2PE or 3PE measurements agree within error. However, they diverge below this temperature due to the difficulty in determining $T_2$ accurately using 3PE measurements when $T_2\gg\tau$. Instead, the long tail of the $T_{W}$-dependent decay visible in Figs. 4c and 4d is sensitive to spectral-diffusion processes that occur on timescales much longer than $T_2$. For this reason we rely solely on the 2PE fit to determine $T_{2}$ at the lowest spin-bath temperatures, and we observe a maximum value of $T_2 = 1.3\pm0.1$ ms at $T_B=23$ mK.

\section{Analysis}

Figures 5a and 5b illustrate the large decrease in $\Gamma_{\mathrm{SD}}$ and $R$ with decreasing temperature for both the Er and Yb spins. We attribute this to a reduction in the electron-spin flip-flop rate as a consequence of thermal polarisation of the spin populations. Indeed, our a-priori neglect of spin-lattice contribution in our SD analysis (Sup. Mat. Section V) is justified by the rapid spin-flip rates observed across the studied temperature range. For instance, $R^{Er}$ is orders-of-magnitude greater than the spin-lattice relaxation rate ${T_{1}}^{-1} = 0.01\,{\textrm{s}}^{-1}$ estimated from previous measurements of spin-lattice relaxation in $\text{Er}\text{:CaWO}_{4}$ at 7.88 GHz and 500 mK \cite{Dantec2021}, given the $g_{S}^{3} B_{0}^{5}$ scaling presented in Equation 4. 

While flip-flops within the Er electronic-spin bath are clearly the dominant source of SD at high temperature, the Yb electron-spins contribute most to SD at $T_B=23$ mK. This occurs because the Yb spins are less thermally polarised as a consequence of their smaller g-factor, even though the Yb spin-bath is less dense than the Er bath. Moreover, the total experimental contribution to SD from both the Er and Yb spins at the lowest temperature is significantly greater than the expected contribution shown by the dashed-line fits in Figs. 5a \& 5b. 

This discrepancy is also present in the decoherence-rate model presented in Figure 5c. This model is detailed in Section VI of the supplementary materials and takes into account the expected contribution of the temperature-dependent instantaneous diffusion due to pulse-excitation in addition to the dephasing caused by the ${}^{183}$W nuclear spin bath. Here the instantaneous diffusion contribution is temperature-dependent due to the thermal population density of the $m_{I}=|+\nicefrac{3}{2}\rangle$ hyperfine state. Using the method described in Refs \cite{Kurshev1992, Tyryshkin2011} we estimate minimum and maximum contributions of 51 Hz at 23 mK and 630 Hz at 530 mK, respectively. The temperature-independent dephasing caused by the unpolarised ${}^{183}$W nuclear spin bath contributes only 12 Hz; a value derived from the cluster-correlation expansion estimate of $T_2$ determined by Le Dantec et. al. \cite{Dantec2021} for the Er electron-spin transition in $\text{CaWO}_{4}$. 

The remaining discrepancy can have two possible origins. The first is that the uncorrelated-sudden-jump model presented in Section \ref{sec:SD} makes the incorrect assumption that all spin-flips are uncorrelated in this system. Although this is valid for spin-flips driven predominantly by spin-lattice relaxation, as has been the case in previous applications of this model \cite{Bottger2006}, there is an inherent correlation between spins-flips when they are driven by flip-flops. While the derivation of a `correlated-sudden-jump' model which accounts for this discrepancy is beyond the scope of this work, it would be possible to investigate this issue further by simulating flip-flops of the paramagnetic spin baths using the cluster-correlation expansion approach \cite{Yang2008,Yang2009a}.

The second possibility is that one or more additional electron-spin species are contributing significantly to SD at low temperatures, consistent with the observation of additional impurities in the mass-spectrum (Table I). One likely candidate species is $\mathrm{Mn}^{2+}$; the third most abundant impurity in the crystal at a relative concentration of 2.8 ppm. With a g-factor of 2, manganese is weakly polarised at 23 mK and an estimate of the SD contribution were not undertaken here due to the complex electronic level structure in mT fields. \cite{Hempstead1960a}. 

\section{Conclusion}

Here we demonstrated an electron-spin coherence time exceeding 1 millisecond in 20 ppm doped $\text{Er}^{3+}\text{:CaWO}_{4}$, achieved by cooling the electronic spin bath to 23 mK and thereby significantly reducing the electron-spin flip-flop rate of both the Er and Yb impurities. 

Further enhancements in spin-coherence could be achieved by chemically purifying the host matrix or improving the thermal spin-polarisation. For instance, increasing the resonator frequency from 4.4 to 8 gigahertz would significantly improve thermal-polarisation while maintaining compatibility with superconducting microwave circuits that typically operate in the 4-8 GHz band. In this situation one would expect the SD contribution of the Er and Yb spins to be reduced by factors of approximately $10^3$ and $10^2$ respectively, at a spin-bath temperature of 23 mK.

Chemical purification is also an important avenue to pursue because the smaller g-factors of non-erbium impurities will generally lead to significant SD contributions at low temperatures. Indeed, ytterbium alone presents an SD contribution $320$ times greater than erbium at 4.4 GHz and 23 mK. Reducing the Yb concentration to the ppb level observed in chemically purified $\text{CaWO}_{4}$ would reduce this contribution by seven orders of magnitude, assuming the inhomogeneous spin-transition linewidth remains largely unchanged.

Such improvements could result in coherence-times approaching the nuclear-spin-limited $T_2$ of 27 milliseconds \cite{Dantec2021}, while maintaining ppm Er concentrations.

\subsection*{Acknowledgements}
We acknowledge technical support from P. Sénat, D. Duet, P.-F. Orfila, and S. Delprat, and we are grateful for fruitful discussions within the Quantronics group. We acknowledge IARPA and Lincoln Labs for providing the Josephson traveling-wave parametric amplifier.

This project has received funding from the European Union’s Horizon 2020 research and innovation program under Marie Sklodowska-Curie grant agreement no. 765267 (QuSCO) and no. 792727 (SMERC). E.F. acknowledges support from the Agence Nationale de la Recherche (ANR) grant DARKWADOR:ANR-19-CE47-0004. We acknowledge support from the ANR through the Chaire Industrielle NASNIQ under contract ANR-17-CHIN-0001 cofunded by Atos and through the project MIRESPIN under contract ANR-19-CE47-0011 and of the Region Ile-de-France through the DIM SIRTEQ (REIMIC project). We acknowledge support of the AIDAS virtual joint laboratory. R.B.L. was supported by Hong Kong Research Grants Council - General Research Fund (Project 14302121). S.L. was supported by the Impact Postdoctoral Fellowship of CUHK. S.B. thanks the support of the CNRS research infrastructure RENARD (FR 3443)

\newpage

\bibliographystyle{naturemag}
\bibliography{library}

\begin{thebibliography}{10}
\expandafter\ifx\csname url\endcsname\relax
  \def\url#1{\texttt{#1}}\fi
\expandafter\ifx\csname urlprefix\endcsname\relax\def\urlprefix{URL }\fi
\providecommand{\bibinfo}[2]{#2}
\providecommand{\eprint}[2][]{\url{#2}}

\bibitem{Zhong2015a}
\bibinfo{author}{Zhong, M.} \emph{et~al.}
\newblock \bibinfo{title}{{Optically addressable nuclear spins in a solid with
  a six-hour coherence time.}}
\newblock \emph{\bibinfo{journal}{Nature}} \textbf{\bibinfo{volume}{517}},
  \bibinfo{pages}{177--180} (\bibinfo{year}{2015}).
\newblock \urlprefix\url{http://dx.doi.org/10.1038/nature14025}.

\bibitem{Holzapfel2020a}
\bibinfo{author}{Holz{\"{a}}pfel, A.} \emph{et~al.}
\newblock \bibinfo{title}{{Optical storage for 0.53 s in a solid-state atomic
  frequency comb memory using dynamical decoupling}}.
\newblock \emph{\bibinfo{journal}{New Journal of Physics}}
  \textbf{\bibinfo{volume}{22}}, \bibinfo{pages}{063009}
  (\bibinfo{year}{2020}).
\newblock
  \urlprefix\url{https://iopscience.iop.org/article/10.1088/1367-2630/ab8aac
  https://iopscience.iop.org/article/10.1088/1367-2630/ab8aac/meta}.
\newblock \eprint{1910.08009}.

\bibitem{Longdell2005}
\bibinfo{author}{Longdell, J.~J.}, \bibinfo{author}{Fraval, E.},
  \bibinfo{author}{Sellars, M.~J.} \& \bibinfo{author}{Manson, N.~B.}
\newblock \bibinfo{title}{{Stopped Light with Storage Times Greater than One
  Second Using Electromagnetically Induced Transparency in a Solid}}.
\newblock \emph{\bibinfo{journal}{Physical Review Letters}}
  \textbf{\bibinfo{volume}{95}}, \bibinfo{pages}{063601}
  (\bibinfo{year}{2005}).
\newblock
  \urlprefix\url{http://link.aps.org/doi/10.1103/PhysRevLett.95.063601}.

\bibitem{Rancic2017}
\bibinfo{author}{Ran{\v{c}}i{\'{c}}, M.}, \bibinfo{author}{Hedges, M.~P.},
  \bibinfo{author}{Ahlefeldt, R.~L.} \& \bibinfo{author}{Sellars, M.~J.}
\newblock \bibinfo{title}{{Coherence time of over a second in a
  telecom-compatible quantum memory storage material}}.
\newblock \emph{\bibinfo{journal}{Nature Physics}}
  \textbf{\bibinfo{volume}{14}}, \bibinfo{pages}{50--54}
  (\bibinfo{year}{2018}).
\newblock \urlprefix\url{http://www.nature.com/doifinder/10.1038/nphys4254}.

\bibitem{Hedges2010}
\bibinfo{author}{Hedges, M.~P.}, \bibinfo{author}{Longdell, J.~J.},
  \bibinfo{author}{Li, Y.} \& \bibinfo{author}{Sellars, M.~J.}
\newblock \bibinfo{title}{{Efficient quantum memory for light}}.
\newblock \emph{\bibinfo{journal}{Nature}} \textbf{\bibinfo{volume}{465}},
  \bibinfo{pages}{1052--1056} (\bibinfo{year}{2010}).
\newblock \urlprefix\url{http://www.ncbi.nlm.nih.gov/pubmed/20577210
  http://dx.doi.org/10.1038/nature09081}.

\bibitem{Dajczgewand2014}
\bibinfo{author}{Dajczgewand, J.}, \bibinfo{author}{{Le Gou{\"{e}}t}, J.-L.},
  \bibinfo{author}{Louchet-Chauvet, A.} \& \bibinfo{author}{Chaneli{\`{e}}re,
  T.}
\newblock \bibinfo{title}{{Large efficiency at telecom wavelength for optical
  quantum memories}}.
\newblock \emph{\bibinfo{journal}{Optics Letters}}
  \textbf{\bibinfo{volume}{39}}, \bibinfo{pages}{2711} (\bibinfo{year}{2014}).
\newblock
  \urlprefix\url{https://www.osapublishing.org/abstract.cfm?URI=ol-39-9-2711}.

\bibitem{Sabooni2013}
\bibinfo{author}{Sabooni, M.}, \bibinfo{author}{Li, Q.},
  \bibinfo{author}{Kr{\"{o}}ll, S.} \& \bibinfo{author}{Rippe, L.}
\newblock \bibinfo{title}{{Efficient Quantum Memory Using a Weakly Absorbing
  Sample}}.
\newblock \emph{\bibinfo{journal}{Physical Review Letters}}
  \textbf{\bibinfo{volume}{110}}, \bibinfo{pages}{133604}
  (\bibinfo{year}{2013}).
\newblock
  \urlprefix\url{http://link.aps.org/doi/10.1103/PhysRevLett.110.133604}.

\bibitem{Bartholomew2020}
\bibinfo{author}{Bartholomew, J.~G.} \emph{et~al.}
\newblock \bibinfo{title}{{On-chip coherent microwave-to-optical transduction
  mediated by ytterbium in YVO4}}.
\newblock \emph{\bibinfo{journal}{Nature Communications}}
  \textbf{\bibinfo{volume}{11}}, \bibinfo{pages}{1--6} (\bibinfo{year}{2020}).
\newblock \urlprefix\url{https://www.nature.com/articles/s41467-020-16996-x}.
\newblock \eprint{1912.03671}.

\bibitem{Fernandez2019}
\bibinfo{author}{Fernandez-Gonzalvo, X.}, \bibinfo{author}{Horvath, S.~P.},
  \bibinfo{author}{Chen, Y.~H.} \& \bibinfo{author}{Longdell, J.~J.}
\newblock \bibinfo{title}{{Cavity-enhanced Raman heterodyne spectroscopy in
  $\mbox{Er}^{3+}\mbox{:Y}_{2}\mbox{SiO}_{5}$ for microwave to optical signal
  conversion}}.
\newblock \emph{\bibinfo{journal}{Physical Review A}}
  \textbf{\bibinfo{volume}{100}}, \bibinfo{pages}{033807}
  (\bibinfo{year}{2019}).
\newblock \eprint{1712.07735}.

\bibitem{Grezes2016}
\bibinfo{author}{Grezes, C.} \emph{et~al.}
\newblock \bibinfo{title}{{Towards a spin-ensemble quantum memory for
  superconducting qubits}} (\bibinfo{year}{2016}).
\newblock \eprint{1510.06565}.

\bibitem{Sigillito2014}
\bibinfo{author}{Sigillito, A.~J.} \emph{et~al.}
\newblock \bibinfo{title}{{Fast, low-power manipulation of spin ensembles in
  superconducting microresonators}}.
\newblock \emph{\bibinfo{journal}{Applied Physics Letters}}
  \textbf{\bibinfo{volume}{104}}, \bibinfo{pages}{222407}
  (\bibinfo{year}{2014}).
\newblock \urlprefix\url{http://aip.scitation.org/doi/10.1063/1.4881613}.
\newblock \eprint{1403.0018}.

\bibitem{Grezes2015}
\bibinfo{author}{Gr{\`{e}}zes, C.}
\newblock \emph{\bibinfo{title}{{Towards a spin ensemble quantum memory for
  superconducting qubits}}}.
\newblock \bibinfo{type}{Phd thesis}, \bibinfo{school}{Universite Paris-Saclay}
  (\bibinfo{year}{2015}).

\bibitem{Ranjan2020}
\bibinfo{author}{Ranjan, V.} \emph{et~al.}
\newblock \bibinfo{title}{{Electron spin resonance spectroscopy with femtoliter
  detection volume}}.
\newblock \emph{\bibinfo{journal}{Applied Physics Letters}}
  \textbf{\bibinfo{volume}{116}}, \bibinfo{pages}{184002}
  (\bibinfo{year}{2020}).
\newblock \urlprefix\url{https://aip.scitation.org/doi/abs/10.1063/5.0004322}.

\bibitem{Probst2015}
\bibinfo{author}{Probst, S.}, \bibinfo{author}{Rotzinger, H.},
  \bibinfo{author}{Ustinov, A.~V.} \& \bibinfo{author}{Bushev, P.~A.}
\newblock \bibinfo{title}{{Microwave multimode memory with an erbium spin
  ensemble}}.
\newblock \emph{\bibinfo{journal}{Physical Review B - Condensed Matter and
  Materials Physics}} \textbf{\bibinfo{volume}{92}}, \bibinfo{pages}{014421}
  (\bibinfo{year}{2015}).
\newblock
  \urlprefix\url{https://journals.aps.org/prb/abstract/10.1103/PhysRevB.92.014421}.
\newblock \eprint{1501.01499}.

\bibitem{Dold2020}
\bibinfo{author}{Dold, G.}
\newblock \emph{\bibinfo{title}{{milliKelvin ESR of rare-earth doped crystals
  using superconducting resonators}}}.
\newblock \bibinfo{type}{Phd thesis}, \bibinfo{school}{University College
  London} (\bibinfo{year}{2020}).

\bibitem{Li2020}
\bibinfo{author}{Li, P.~Y.} \emph{et~al.}
\newblock \bibinfo{title}{{Hyperfine Structure and Coherent Dynamics of
  Rare-Earth Spins Explored with Electron-Nuclear Double Resonance at Subkelvin
  Temperatures}}.
\newblock \emph{\bibinfo{journal}{Physical Review Applied}}
  \textbf{\bibinfo{volume}{13}}, \bibinfo{pages}{024080}
  (\bibinfo{year}{2020}).
\newblock \eprint{1910.12351}.

\bibitem{Rakonjac2020}
\bibinfo{author}{Rakonjac, J.~V.}, \bibinfo{author}{Chen, Y.-H.},
  \bibinfo{author}{Horvath, S.~P.} \& \bibinfo{author}{Longdell, J.~J.}
\newblock \bibinfo{title}{{Long spin coherence times in the ground state and in
  an optically excited state of $^{167}$Er$^{3+}$:Y$_2$SiO$_5$ at zero magnetic
  fields}}.
\newblock \emph{\bibinfo{journal}{Physical Review B}} \bibinfo{pages}{184430}
  (\bibinfo{year}{2020}).
\newblock
  \urlprefix\url{https://journals.aps.org/prb/abstract/10.1103/PhysRevB.101.184430}.

\bibitem{Ortu2018}
\bibinfo{author}{Ortu, A.} \emph{et~al.}
\newblock \bibinfo{title}{{Simultaneous coherence enhancement of optical and
  microwave transitions in solid-state electronic spins}}.
\newblock \emph{\bibinfo{journal}{Nature Materials 2018 17:8}}
  \textbf{\bibinfo{volume}{17}}, \bibinfo{pages}{671--675}
  (\bibinfo{year}{2018}).
\newblock \urlprefix\url{https://www.nature.com/articles/s41563-018-0138-x}.
\newblock \eprint{1712.08615}.

\bibitem{Dantec2021}
\bibinfo{author}{Dantec, M.~L.} \emph{et~al.}
\newblock \bibinfo{title}{{Twenty-three millisecond electron spin coherence of
  erbium ions in a natural-abundance crystal}}.
\newblock \emph{\bibinfo{journal}{Science Advances}}
  \textbf{\bibinfo{volume}{7}}, \bibinfo{pages}{9786} (\bibinfo{year}{2021}).
\newblock
  \urlprefix\url{https://www.science.org/doi/full/10.1126/sciadv.abj9786}.
\newblock \eprint{2106.14974}.

\bibitem{Kanai2021}
\bibinfo{author}{Kanai, S.} \emph{et~al.}
\newblock \bibinfo{title}{{Generalized scaling of spin qubit coherence in over
  12,000 host materials}}  (\bibinfo{year}{2021}).
\newblock \urlprefix\url{https://arxiv.org/abs/2102.02986v1}.
\newblock \eprint{2102.02986}.

\bibitem{Herzog1956}
\bibinfo{author}{Herzog, B.} \& \bibinfo{author}{Hahn, E.~L.}
\newblock \bibinfo{title}{{Transient nuclear induction and double nuclear
  resonance in solids}}.
\newblock \emph{\bibinfo{journal}{Physical Review}}
  \textbf{\bibinfo{volume}{103}}, \bibinfo{pages}{148--166}
  (\bibinfo{year}{1956}).

\bibitem{Ferrenti2020}
\bibinfo{author}{Ferrenti, A.~M.}, \bibinfo{author}{de~Leon, N.~P.},
  \bibinfo{author}{Thompson, J.~D.} \& \bibinfo{author}{Cava, R.~J.}
\newblock \bibinfo{title}{{Identifying candidate hosts for quantum defects via
  data mining}}.
\newblock \emph{\bibinfo{journal}{npj Computational Materials}}
  \textbf{\bibinfo{volume}{6}}, \bibinfo{pages}{1--6} (\bibinfo{year}{2020}).
\newblock \urlprefix\url{https://www.nature.com/articles/s41524-020-00391-7}.
\newblock \eprint{2006.14128}.

\bibitem{Kiel1970}
\bibinfo{author}{Kiel, A.} \& \bibinfo{author}{Mims, W.~B.}
\newblock \bibinfo{title}{{Electric-Field-Induced g Shifts for Loose Yb Ions in
  Three Scheelite Lattices}}.
\newblock \emph{\bibinfo{journal}{Physical Review B}}
  \textbf{\bibinfo{volume}{1}}, \bibinfo{pages}{2935--2944}
  (\bibinfo{year}{1970}).
\newblock \urlprefix\url{https://link.aps.org/doi/10.1103/PhysRevB.1.2935}.

\bibitem{Bertaina2007}
\bibinfo{author}{Bertaina, S.} \emph{et~al.}
\newblock \bibinfo{title}{{Rare-earth solid-state qubits}}.
\newblock \emph{\bibinfo{journal}{Nature Nanotechnology}}
  \textbf{\bibinfo{volume}{2}}, \bibinfo{pages}{39--42} (\bibinfo{year}{2007}).
\newblock \urlprefix\url{http://www.nature.com/doifinder/10.1038/nnano.2006.174
  http://www.nature.com/articles/nnano.2006.174}.

\bibitem{Guillot-Noel2006b}
\bibinfo{author}{Guillot-No{\"{e}}l, O.} \emph{et~al.}
\newblock \bibinfo{title}{{Hyperfine interaction of $\text{Er}^{3+}$ ions in
  $\text{Y}{}_{2}\text{:SiO}{}_{5}$ : An electron paramagnetic resonance
  spectroscopy study}}.
\newblock \emph{\bibinfo{journal}{Physical Review B}}
  \textbf{\bibinfo{volume}{74}}, \bibinfo{pages}{214409}
  (\bibinfo{year}{2006}).
\newblock \urlprefix\url{http://link.aps.org/doi/10.1103/PhysRevB.74.214409}.

\bibitem{Macklin2015}
\bibinfo{author}{Macklin, C.} \emph{et~al.}
\newblock \bibinfo{title}{{A near-quantum-limited Josephson traveling-wave
  parametric amplifier}}.
\newblock \emph{\bibinfo{journal}{Science}} \textbf{\bibinfo{volume}{350}},
  \bibinfo{pages}{307--310} (\bibinfo{year}{2015}).

\bibitem{Probst2017}
\bibinfo{author}{Probst, S.} \emph{et~al.}
\newblock \bibinfo{title}{{Inductive-detection electron-spin resonance
  spectroscopy with $65 spins/ \sqrt{Hz}$ sensitivity}}.
\newblock \emph{\bibinfo{journal}{Applied Physics Letters}}
  \textbf{\bibinfo{volume}{111}}, \bibinfo{pages}{202604}
  (\bibinfo{year}{2017}).
\newblock \urlprefix\url{http://aip.scitation.org/doi/10.1063/1.5002540}.
\newblock \eprint{1708.09287}.

\bibitem{Bienfait2015a}
\bibinfo{author}{Bienfait, A.} \emph{et~al.}
\newblock \bibinfo{title}{{Reaching the quantum limit of sensitivity in
  electron spin resonance}}.
\newblock \emph{\bibinfo{journal}{Nature nanotechnology}}
  \textbf{\bibinfo{volume}{11}}, \bibinfo{pages}{253--257}
  (\bibinfo{year}{2015}).
\newblock
  \urlprefix\url{http://www.nature.com/doifinder/10.1038/nnano.2015.282}.
\newblock \eprint{1507.06831}.

\bibitem{Song2009}
\bibinfo{author}{Song, C.} \emph{et~al.}
\newblock \bibinfo{title}{{Microwave response of vortices in superconducting
  thin films of Re and Al}}.
\newblock \emph{\bibinfo{journal}{Physical Review B - Condensed Matter and
  Materials Physics}} \textbf{\bibinfo{volume}{79}}, \bibinfo{pages}{174512}
  (\bibinfo{year}{2009}).
\newblock
  \urlprefix\url{https://journals.aps.org/prb/abstract/10.1103/PhysRevB.79.174512}.
\newblock \eprint{0812.3645}.

\bibitem{Probst2013a}
\bibinfo{author}{Probst, S.} \emph{et~al.}
\newblock \bibinfo{title}{{Anisotropic rare-earth spin ensemble strongly
  coupled to a superconducting resonator.}}
\newblock \emph{\bibinfo{journal}{Physical review letters}}
  \textbf{\bibinfo{volume}{110}}, \bibinfo{pages}{157001}
  (\bibinfo{year}{2013}).
\newblock
  \urlprefix\url{https://link.aps.org/doi/10.1103/PhysRevLett.110.157001
  https://journals.aps.org/prl/abstract/10.1103/PhysRevLett.110.157001}.

\bibitem{Kubo2010}
\bibinfo{author}{Kubo, Y.} \emph{et~al.}
\newblock \bibinfo{title}{{Strong Coupling of a Spin Ensemble to a
  Superconducting Resonator}}.
\newblock \emph{\bibinfo{journal}{Physical Review Letters}}
  \textbf{\bibinfo{volume}{105}}, \bibinfo{pages}{140502}
  (\bibinfo{year}{2010}).
\newblock
  \urlprefix\url{https://link.aps.org/doi/10.1103/PhysRevLett.105.140502}.

\bibitem{Mims1961}
\bibinfo{author}{Mims, W.~B.}, \bibinfo{author}{Nassau, K.} \&
  \bibinfo{author}{McGee, J.~D.}
\newblock \bibinfo{title}{{Spectral Diffusion in Electron Resonance Lines}}.
\newblock \emph{\bibinfo{journal}{Physical Review}}
  \textbf{\bibinfo{volume}{123}}, \bibinfo{pages}{2059} (\bibinfo{year}{1961}).
\newblock
  \urlprefix\url{https://journals.aps.org/pr/abstract/10.1103/PhysRev.123.2059}.

\bibitem{Hu1974}
\bibinfo{author}{Hu, P.} \& \bibinfo{author}{Hartmann, S.~R.}
\newblock \bibinfo{title}{{Theory of spectral diffusion decay using an
  uncorrelated-sudden-jump model}}.
\newblock \emph{\bibinfo{journal}{Physical Review B}}
  \textbf{\bibinfo{volume}{9}}, \bibinfo{pages}{1} (\bibinfo{year}{1974}).
\newblock
  \urlprefix\url{https://journals.aps.org/prb/abstract/10.1103/PhysRevB.9.1}.

\bibitem{Bottger2006}
\bibinfo{author}{B{\"{o}}ttger, T.}, \bibinfo{author}{Thiel, C.~W.},
  \bibinfo{author}{Sun, Y.} \& \bibinfo{author}{Cone, R.~L.}
\newblock \bibinfo{title}{{Optical decoherence and spectral diffusion at 1.5
  $\mu$m in $\mbox{Er}^{3+}\mbox{:Y}_{2}\mbox{SiO}_{5}$ versus magnetic field,
  temperature, and $\mbox{Er}^{3+}$ concentration}}.
\newblock \emph{\bibinfo{journal}{Physical Review B}}
  \textbf{\bibinfo{volume}{73}}, \bibinfo{pages}{075101}
  (\bibinfo{year}{2006}).
\newblock \urlprefix\url{http://link.aps.org/doi/10.1103/PhysRevB.73.075101}.

\bibitem{Bai1989}
\bibinfo{author}{Bai, Y.~S.} \& \bibinfo{author}{Fayer, M.~D.}
\newblock \bibinfo{title}{{Time scales and optical dephasing measurements:
  Investigation of dynamics in complex systems}}.
\newblock \emph{\bibinfo{journal}{Physical Review B}}
  \textbf{\bibinfo{volume}{39}}, \bibinfo{pages}{11066} (\bibinfo{year}{1989}).
\newblock
  \urlprefix\url{https://journals.aps.org/prb/abstract/10.1103/PhysRevB.39.11066}.

\bibitem{Probst2020a}
\bibinfo{author}{Probst, S.} \emph{et~al.}
\newblock \bibinfo{title}{{Hyperfine spectroscopy in a quantum-limited
  spectrometer}}.
\newblock \emph{\bibinfo{journal}{Magnetic Resonance}}
  \textbf{\bibinfo{volume}{1}}, \bibinfo{pages}{315--330}
  (\bibinfo{year}{2020}).
\newblock \eprint{2001.04854}.

\bibitem{An}
\bibinfo{author}{Angeletti, B.} \& \bibinfo{author}{Ambrosi, J.~P.}
\newblock \bibinfo{title}{{European Centre for Environmental Geoscience
  Research and Teaching}} (\bibinfo{year}{2020}).
\newblock
  \urlprefix\url{https://www.cerege.fr/en/analytic-facilities/la-icp-ms-platform-elemental-chemistry}.

\bibitem{Dantec2022}
\bibinfo{author}{Dantec, M.~L.}
\newblock \emph{\bibinfo{title}{{Electron spin dynamics of erbium ions in
  scheelite crystals , probed with superconducting resonators at millikelvin
  temperatures}}}.
\newblock Ph.D. thesis, \bibinfo{school}{Universit{\'{e}} Paris-Saclay}
  (\bibinfo{year}{2022}).
\newblock \urlprefix\url{https://tel.archives-ouvertes.fr/tel-03579857}.

\bibitem{Car2019}
\bibinfo{author}{Car, B.}, \bibinfo{author}{Veissier, L.},
  \bibinfo{author}{Louchet-Chauvet, A.}, \bibinfo{author}{{Le Gou{\"{e}}t},
  J.~L.} \& \bibinfo{author}{Chaneli{\`{e}}re, T.}
\newblock \bibinfo{title}{{Optical study of the anisotropic erbium spin
  flip-flop dynamics}}.
\newblock \emph{\bibinfo{journal}{Physical Review B}}
  \textbf{\bibinfo{volume}{100}}, \bibinfo{pages}{165107}
  (\bibinfo{year}{2019}).
\newblock
  \urlprefix\url{https://journals.aps.org/prb/abstract/10.1103/PhysRevB.100.165107}.
\newblock \eprint{1811.10285}.

\bibitem{Mims1966}
\bibinfo{author}{Mims, W.~B.} \& \bibinfo{author}{Gillen, R.}
\newblock \bibinfo{title}{{Broadening of paramagnetic-resonance lines by
  internal electric fields}}.
\newblock \emph{\bibinfo{journal}{Physical Review}}
  \textbf{\bibinfo{volume}{148}}, \bibinfo{pages}{438--443}
  (\bibinfo{year}{1966}).
\newblock
  \urlprefix\url{https://journals.aps.org/pr/abstract/10.1103/PhysRev.148.438}.

\bibitem{Baibekov2014}
\bibinfo{author}{Baibekov, E.~I.} \emph{et~al.}
\newblock \bibinfo{title}{{Broadening of paramagnetic resonance lines by
  charged point defects in neodymium-doped scheelites}}.
\newblock \emph{\bibinfo{journal}{Optics and Spectroscopy}}
  \textbf{\bibinfo{volume}{116}}, \bibinfo{pages}{661--666}
  (\bibinfo{year}{2014}).
\newblock \urlprefix\url{http://link.springer.com/10.1134/S0030400X1405004X}.

\bibitem{Ranjan2013}
\bibinfo{author}{Ranjan, V.} \emph{et~al.}
\newblock \bibinfo{title}{{Probing Dynamics of an Electron-Spin Ensemble via a
  Superconducting Resonator}}.
\newblock \emph{\bibinfo{journal}{Physical Review Letters}}
  \textbf{\bibinfo{volume}{110}}, \bibinfo{pages}{067004}
  (\bibinfo{year}{2013}).
\newblock
  \urlprefix\url{https://link.aps.org/doi/10.1103/PhysRevLett.110.067004}.
\newblock \eprint{1208.5473}.

\bibitem{Zech2019}
\bibinfo{author}{Zech, M.}, \bibinfo{author}{Schoebel, J.} \&
  \bibinfo{author}{Pickert, T.}
\newblock \bibinfo{title}{{Stick-slip drive, especially pieze-actuated inertial
  drive}} (\bibinfo{year}{2019}).
\newblock \urlprefix\url{https://patents.google.com/patent/US10505470B2/en}.

\bibitem{Kurshev1992}
\bibinfo{author}{Kurshev, V.~V.} \& \bibinfo{author}{Ichikawa, T.}
\newblock \bibinfo{title}{{Effect of spin flip-flop on electron-spin-echo decay
  due to instantaneous diffusion}}.
\newblock \emph{\bibinfo{journal}{Journal of Magnetic Resonance (1969)}}
  \textbf{\bibinfo{volume}{96}}, \bibinfo{pages}{563--573}
  (\bibinfo{year}{1992}).

\bibitem{Tyryshkin2011}
\bibinfo{author}{Tyryshkin, A.~M.} \emph{et~al.}
\newblock \bibinfo{title}{{Electron spin coherence exceeding seconds in
  high-purity silicon}}.
\newblock \emph{\bibinfo{journal}{Nature Materials 2011 11:2}}
  \textbf{\bibinfo{volume}{11}}, \bibinfo{pages}{143--147}
  (\bibinfo{year}{2011}).
\newblock \urlprefix\url{https://www.nature.com/articles/nmat3182}.

\bibitem{Yang2008}
\bibinfo{author}{Yang, W.} \& \bibinfo{author}{Liu, R.~B.}
\newblock \bibinfo{title}{{Quantum many-body theory of qubit decoherence in a
  finite-size spin bath}}.
\newblock \emph{\bibinfo{journal}{Physical Review B - Condensed Matter and
  Materials Physics}} \textbf{\bibinfo{volume}{78}}, \bibinfo{pages}{085315}
  (\bibinfo{year}{2008}).
\newblock \eprint{0806.0098}.

\bibitem{Yang2009a}
\bibinfo{author}{Yang, W.} \& \bibinfo{author}{Liu, R.~B.}
\newblock \bibinfo{title}{{Quantum many-body theory of qubit decoherence in a
  finite-size spin bath. II. Ensemble dynamics}}.
\newblock \emph{\bibinfo{journal}{Physical Review B - Condensed Matter and
  Materials Physics}} \textbf{\bibinfo{volume}{79}}, \bibinfo{pages}{115320}
  (\bibinfo{year}{2009}).
\newblock \eprint{0902.3055}.

\bibitem{Hempstead1960a}
\bibinfo{author}{Hempstead, C.~F.} \& \bibinfo{author}{Bowers, K.~D.}
\newblock \bibinfo{title}{{Paramagnetic resonance of impurities in CaWO4. I.
  Two S-state ions}}.
\newblock \emph{\bibinfo{journal}{Physical Review}}
  \textbf{\bibinfo{volume}{118}}, \bibinfo{pages}{131--134}
  (\bibinfo{year}{1960}).

\end{thebibliography}

\end{document}


\title{Supplementary Materials}
\maketitle

\section*{Section I: Measurements of Instantaneous Diffusion}

\begin{figure}[H]
\begin{centering}
\includegraphics[scale=0.5]{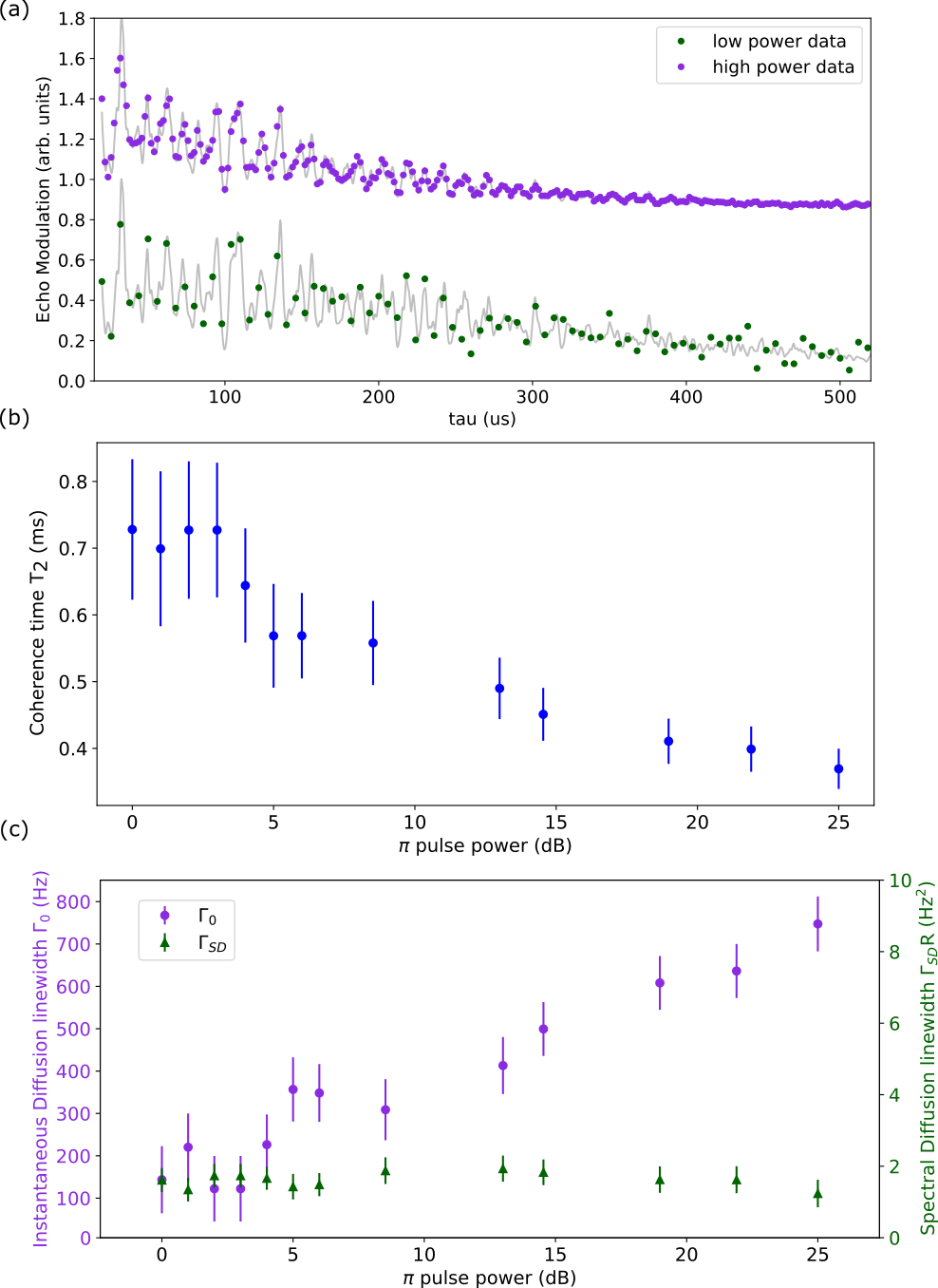}
\par\end{centering}
\caption{\label{fig:ID_plot}Two-pulse echo data recorded on the $I=0$ ESR
transition of $\mathrm{Er}^{3+}$ at 43mT. (a) 2PE time traces taken
with high and low microwave power (0 and 25 dB). The two time traces
shown here represent the average of 120 and 1000 echo measurements,
for the top and bottom traces respectively. (b) 2PE coherence time
as a function of applied microwave power. (c) Instantaneous and spectral
diffusion contributions to the 2PE decay rate as a function of applied
microwave power.}

\end{figure}

Instantaneous diffusion is often not an intrinsic decoherence mechanism
of the material, and is usually the result of applied microwave pulses
which perturb the spins and thereby induce dephasing via the magnetic-dipole
interaction between the excited spins. This phenomenon is often a
significant source of dephasing for ESR transitions with very high
spin densities and its contribution can be quantified by studying
the coherence time of two-pulse-echoes (2PE) as a function of applied
microwave pulse power. For the Er $I=0$ transition in particular,
$T_{2}$ changes drastically with applied power. This can be seen
in Figure \ref{fig:ID_plot} (a) \& (b). 

Although it is possible to achieve longer coherence times in the presence
of ID by reducing the applied microwave pulse power, the longest $T_{2}$
time achieved here on the Er $I=0$ transition is only 720$\mu s$,
as shown in Figure 1b. This is considerably shorter than the 1.3ms
measured on the $m_{I}=|+\nicefrac{3}{2}\rangle$ sub-ensemble at
the same spin-bath temperature. Figure 1c shows the relative contribution
of instantaneous and spectral diffusion in this decay rate when the
2PE curves are fitted using a phenomenological spectral diffusion
model based on Eq. (1) of the main text:
\begin{alignat*}{1}
\text{A}_{e}= & \text{V}_{eseem}\left(\tau,T_{W}\right)\cdot\text{A}_{0}\exp\left(-2\pi\tau\left[\Gamma_{0}+\tau\Gamma_{\text{SD}}R/2\right]\right)
\end{alignat*}
Here $\Gamma_{0}$ demonstrates a strong dependence on pulse-power
(Fig. 1c) which is consistent with ID caused by pulse-driven inter-spin
interactions. However, we do not observe the expected $\sin^{2}\theta$
dependence with respect to applied-pulse angle $\theta$ \cite{Kurshev1992,Tyryshkin2011}.
This is likely because of the spatial-inhomogeneity of the spin-resonator
coupling (and hence inhomogeneity of the Rabi angle) for the micro-resonator
geometry used here. Moreover, the spin-excitation volume also changes
with applied microwave power which can be a confounding factor if
the magnetic interactions are sufficiently long-range.

\newpage

\section*{Section II: Magnetic Rotation Spectrum}

\begin{figure}[H]
\centering{}\includegraphics[scale=0.5]{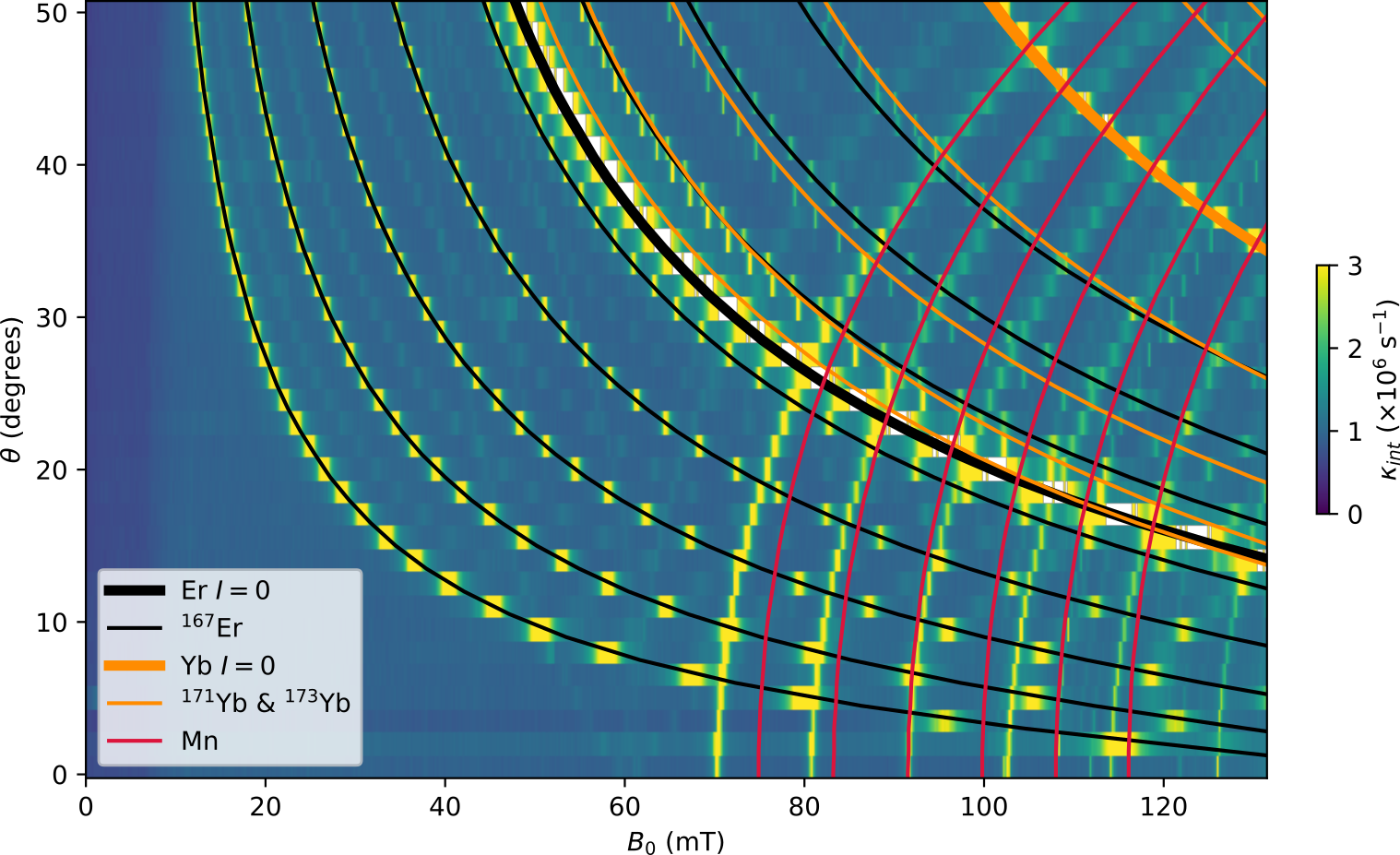}\caption{\label{fig:rotation-pattern}MRS spectrum comprising 34 continuous-wave
ESR transmission spectra, each recorded as function of magnetic-field
with 1.5 degree increments of field angle between each spectrum, with
$\theta=0$ corresponding to $B_{0}\parallel c$. Spin transitions
can be seen here by the decrease in resonator Q caused by the spin-resonator
coupling. \textit{Coloured solid lines}: Theoretical spectra of $\text{Er}^{3+}$,
$\text{Yb}^{3+}$ and $\text{Mn}^{2+}$ based on spin-Hamiltonian
parameters of Refs \cite{Antipin1968,Kiel1970,Hempstead1960a} respectively.
For isotopes with non-zero nuclear spin $\left(^{167}\text{Er},{}^{171}\text{Yb}\,\text{and}\,{}^{173}\text{Yb}\right)$
only the nuclear-spin preserving $\Delta m_{I}=0$ transitions are
shown. Due to spin-state mixing, however, some weak $\Delta m_{I}=\pm1$
transitions can be identified in the MRS spectrum above. }
\end{figure}

Magnetic Rotation Spectroscopy (MRS) is a centuries old technique\cite{Righi1898},
and extremely useful for identifying particular paramagnetic species
in complex and dense ESR spectra\cite{McCarthy1992}. Here we rotate
the magnetic field in the $a-c$ plane of the crystal and record ESR
transmission spectra every 1.5 degrees. Several paramagnetic impurities
are resolved in this way, shown by the MRS spectrum in Figure \ref{fig:rotation-pattern}.
Here we identify $\text{Mn}^{2+}$ and $\text{Yb}^{3+}$ impurities
in addition to the known $\text{Er}^{3+}$ impurities, using known
Zeeman-Hamiltonian parameters, in particular g-tensors \cite{Kiel1970,Hempstead1960a}. 

In the MSR spectrum the $\text{Mn}^{2+}$ and $\text{Yb}^{3+}$ impurities
present six and eight ESR lines respectively. The 6 $\text{Mn}^{2+}$
lines are attributed to the $^{55}\text{Mn}$ isotope, with nuclear
$I=\nicefrac{3}{2}$ and 100\% natural abundance. For the $\text{Yb}^{3+}$
impurities, two lines are attributed to $^{171}\text{Yb}$ ($I=\nicefrac{1}{2},$
14\% nat. abun.), five lines are attributed to $^{173}\text{Yb}$
($I=\nicefrac{5}{2},$ 16\% nat. abun.) and the last line is a mixture
of all the even isotopes between $^{168}\text{Yb}$ and $^{176}\text{Yb}$
inclusively ($I=0,$ 70\% total nat. abun.). The concentrations of
$\text{Yb}^{3+}$impurities relative to $\text{Er}^{3+}$ can easily
be determined from fits of the dominant $I=0$ anti-crossing (refer
to Section IV below). Estimates of the $\text{Mn}^{2+}$ concentration
were not attempted, however, due to its more complex electronic level
structure.

\newpage

\section*{Section III: Fitting the Transmission Spectra}

\begin{figure}[H]
\centering{}\includegraphics[scale=0.3]{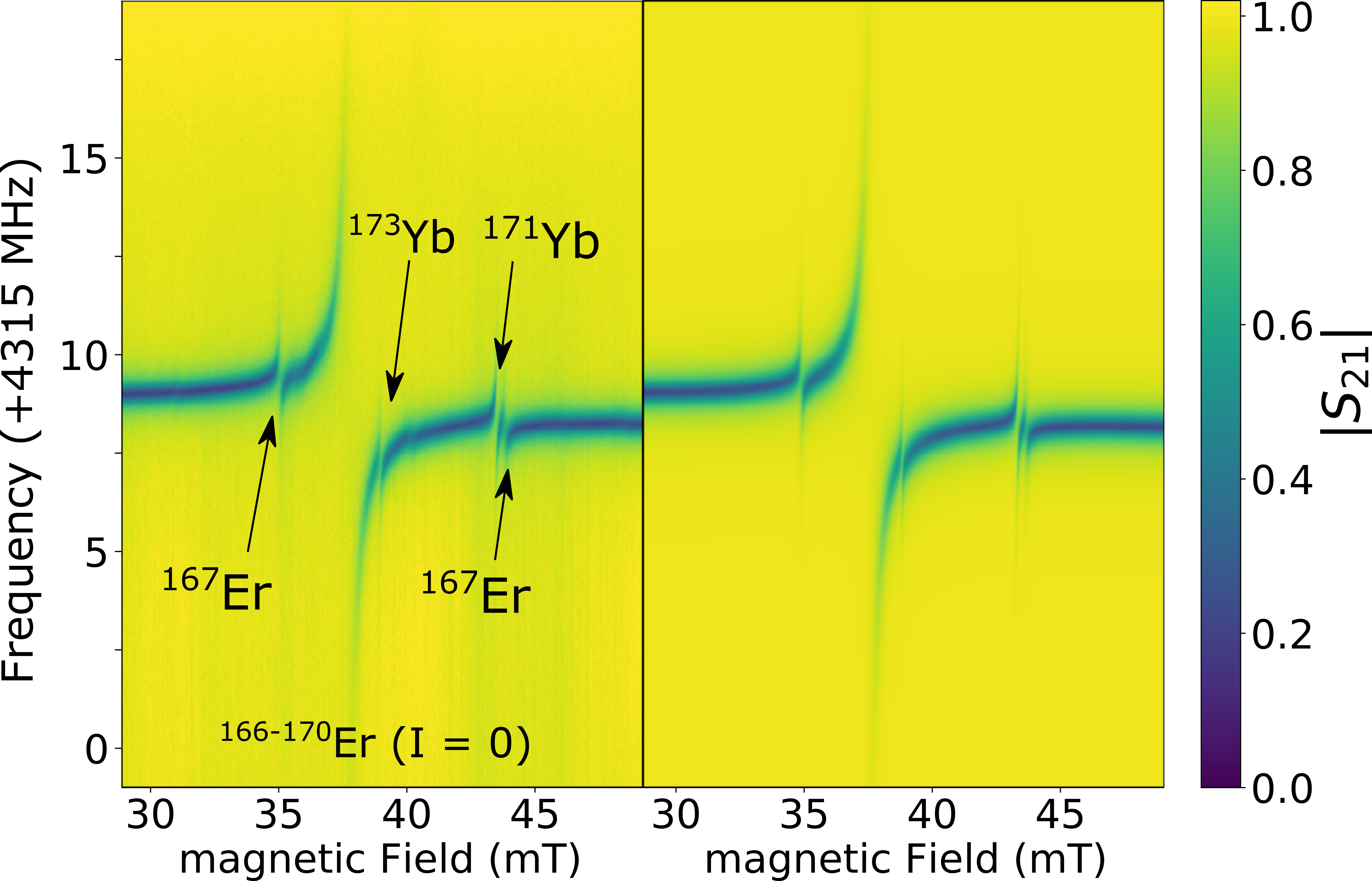}\caption{\label{fig:fitting-S21}\textit{Left}: A sub-section of the transmission
spectrum presented in Fig. 2 of the main text, focusing on the transitions
studied in Fig 3. of the main text. Five avoided level crossingss
are observed in this sub-spectrum, labeled according to atomic and
isotopic composition. \textit{Right}: Fit to the transmission spectrum
using Eq. (5) of the main text, assuming known Hamiltonian parameters
of $\text{Er}^{3+}$ and $\text{Yb}^{3+}$ \cite{Antipin1968,Kiel1970}.}
\end{figure}

Equation (5) of the main text can be used to independently fit both
the spin-resonator coupling $g_{ens}$ and the spin-transition linewidth
$\Gamma$ of several paramagnetic spin species. If the resonator parameters
$\kappa_{c},\,\kappa_{i}(B_{0})$ and the spin-transition frequency
$\omega_{s}(B_{0})$ are well known, then this fit can yield very
accurate results. Figure \ref{fig:fitting-S21} presents a side-by-side
example of measured and fitted transmission spectra recorded at 100mK
as a function of microwave frequency $\omega_{r}/2\pi$ and applied
magnetic field $B_{0}$. This fit is obtained by least-squares minimisation
of $S_{21}$ in the two-dimensional plane shown in Fig. \ref{fig:fitting-S21}.
Here the resonator frequency $\omega_{r}$ is assumed to decrease
quadratically with increasing magnetic field and the quadratic components
are determined by a least-squares fit of $S_{21}$ using only sections
of the ESR spectrum that are devoid of spin transitions (see Fig.
2 of the main text). In this way there are only ten free parameters
required to fit the spectrum shown in Figure \ref{fig:fitting-S21}
below, comprising of the $(g_{ens},\Gamma)$ pairs presented in Table
\ref{tab:coupling-strengths}. To obtain accurate values for these
remaining ten parameters, and especially those of the smaller transitions,
the fit proceeds in two steps:

1) the Er $I=0$ is fitted alone because its large anti-crossing distorts
the spectrum at the location of other nearby anti-crossings. 

2) A simultaneous fitting of the other four avoided level crossings
is performed with the two Er $I=0$ parameters ($g_{ens},\Gamma)$
held fixed. 

From the fit we extract the coupling parameters for the five avoided
level crossings between 30-50mT, which are tabulated Table \ref{tab:coupling-strengths}.
Note that large differences in population between the Er $I=0$ and
the other four ensembles presented a challenge when fitting the smaller
anti-crossings in the spectrum. 

\begin{table}[H]
\begin{centering}
\begin{tabular}{|c|c|c|}
\hline 
Transition & $g_{ens}$(kHz) & $\Gamma$(MHz)\tabularnewline
\hline 
\hline 
$^{167}\text{Er}$ $|+\nicefrac{1}{2}\rangle$ & $\quad2150\pm5\quad$ & $\quad16.3\pm0.1$\tabularnewline
\hline 
$^{166-172}\text{Er}$ $I=0\quad$ & $14135\pm5$ & $36.2\pm0.1$\tabularnewline
\hline 
$^{173}\text{Yb}$ $|-\nicefrac{5}{2}\rangle$ & $1215\pm5$ & $5.9\pm0.1$\tabularnewline
\hline 
$^{171}\text{Yb}$ $|-\nicefrac{1}{2}\rangle$ & $1465\pm5$ & $\quad3.62\pm0.05$\tabularnewline
\hline 
$^{167}\text{Er}$ $|+\nicefrac{3}{2}\rangle$ & $2105\pm5$ & $22.3\pm0.2$\tabularnewline
\hline 
\end{tabular}
\par\end{centering}
\caption{\label{tab:coupling-strengths}Spin-resonator couplings and inhomogeneous
linewidths determined at 100mK using the fit presented in Fig \ref{fig:fitting-S21}.
Transitions are tabulated in order of increasing magnetic field at
the avoided level crossing point. Uncertainties represent the Standard
Error in the fit.}
\end{table}

In this way $g_{ens}$ and $\Gamma$ can be measured as a function
of mixing-chamber temperature, by recording the transmission spectrum
at different temperatures. For the three Er transitions tabulated
above, the resulting $g_{ens}$ and $\Gamma$ values in the range
of 10-550mK values are presented in Fig. 3 of the main text.

\section*{Section IV: Estimating Spin Temperature and REI Impurity Concentrations}

To elucidate the temperature of the electron-spin bath $T_{B}$ in
Fig. 3 of the main text, we compare the fitted ensemble-coupling as
a function of temperature $g_{ens}(T)$ with the theoretical value
of $g_{ens}(T)$ expected from Boltzmann thermal statistics. In particular,
this requires a numerical determination of the spin-resonator coupling
$g$ within the volume of the calcium tungstate crystal (sub-sections
1 \& 2) followed by a theoretical treatment of the temperature dependence
(sub-section 3). This method further yields a direct method for determining
the $\left[\text{Er}^{3+}\right]$ impurity concentration, and by
extension the $\left[\text{Yb}^{3+}\right]$, which can be directly
compared with the known impurity concentrations obtained from mass-spectrometry
results presented in Table I of the main text.

\subsection*{Subsection IV.1: Determining the spin-resonator coupling constant
`g'}

The coupling between a single spin and a harmonic oscillator follows
the Jaynes-Cummings Hamiltonian \cite{Haikka2017}:

\begin{alignat*}{1}
H & =\hbar\omega_{r}\left(\hat{a}^{\dagger}\hat{a}+\frac{1}{2}\right)-\frac{\hbar\omega_{s}(B_{0})}{2}\hat{\sigma}_{z}+\hbar g(\hat{a}\hat{\sigma}_{+}+\hat{a}^{\dagger}\hat{\sigma}_{-})
\end{alignat*}

Here $\hat{a}^{\dagger}$and $\hat{a}$ represent the creation and
annihilation operators of the electromagnetic field (photons) in the
resonator, while $\hat{\sigma}_{+}$ and $\hat{\sigma}_{-}$ represent
raising and lowering operators of the electronic spin moment $S=\nicefrac{1}{2}$.
The other parameters are defined in the main text, and the coupling
constant $g$ is given as follows \cite{Haikka2017}:
\[
g(\boldsymbol{r})=\left|\langle+\nicefrac{1}{2}|\mathbf{\delta B_{1}}(\boldsymbol{r})\cdot\mathbf{\boldsymbol{\gamma}}\cdot\mathbf{S}|-\nicefrac{1}{2}\rangle\right|
\]
Here $\mathbf{\delta B_{1}}(\boldsymbol{r})$ represents the Root-Mean-Square
(RMS) vacuum fluctuations of the magnetic field at the spin's position
$\boldsymbol{r}$, while $\boldsymbol{\gamma}$ is the gyromagnetic-tensor.
For $\text{\ensuremath{\text{Er}^{3+}}:CaWO}_{4}$ and $\text{\ensuremath{\text{Yb}^{3+}}:CaWO}_{4}$
this is given by \cite{Antipin1968,Kiel1970}:

\begin{alignat*}{1}
\gamma_{\text{Er}} & =2\pi\begin{pmatrix}117 & 0 & 0\\
0 & 117 & 0\\
0 & 0 & 17.5
\end{pmatrix}\text{GHz/T}
\end{alignat*}

\begin{alignat*}{1}
\gamma_{\mathrm{Yb}} & =2\pi\begin{pmatrix}55.1 & 0 & 0\\
0 & 55.1 & 0\\
0 & 0 & 14.7
\end{pmatrix}\text{GHz/T}
\end{alignat*}

\begin{figure}[H]
\centering{}\includegraphics[scale=0.8]{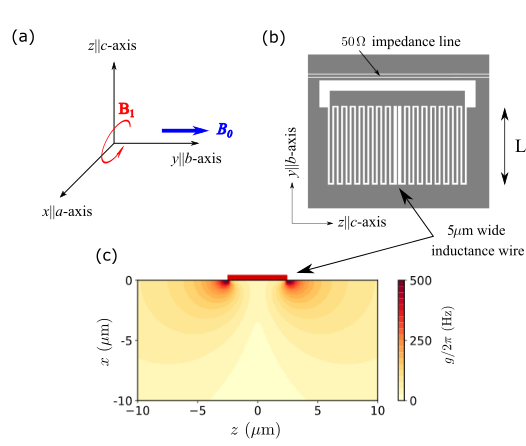}\caption{\label{fig:angles}\textbf{(a)} Coordinate system in the $\left(x,y,z\right)$
frame of the Spin-Hamiltonian and $\left(a,b,c\right)$ frame of the
principle crystal axes. The static magnetic field $B_{0}$ is applied
along the $y$-axis (equivalently to the crystal $b$-axis) while
the oscillating magnetic field $\mathrm{B}_{1}$ only has components
perpendicular to the $y$-axis. \textbf{(b}) Design of the superconducting
resonator patterned on the surface of the $\text{CaWO}_{4}$ crystal,
with respect to crystal axes. Dark grey shows the location of the
niobium thin-film deposited on the surface, exposed crystal is shown
in white\textbf{. }Microwave signals are sent and received via a 50
ohm transmission line, wirebonded on either end to SMP terminals\textbf{
(c) }COMSOL simulation showing the spatial variation of the spin-resonator
coupling $g$ for the Er $I=0$ transitions; the coupling term $g$
is strongly correlated with the strength of the oscillating magnetic
field $\mathrm{B}_{1}$ radiating from the inductive wire. This simulation
is used to calculate the `averaged' gyromagnetic ratio\textbf{ $\tilde{\text{\ensuremath{\gamma}}}$.}}
\end{figure}

The resonator generates an oscillating magnetic field $\mathbf{B_{1}}=\mathbf{\delta B_{1}}(\hat{a}+\hat{a}^{\dagger})$
which is responsible for the spin-resonator interaction and whose
orientation is defined by the geometry of the superconducting inductance-wire
shown in \ref{fig:angles}. In the $\left(x,y,z\right)$ basis, the
vacuum fluctuations of the magnetic field are given by:

\[
\mathbf{\delta B_{1}}(\boldsymbol{r})=\begin{pmatrix}\delta B_{1x}\\
\delta B_{1y}\\
\delta B_{1z}
\end{pmatrix}
\]
for the system studied here, the static magnetic field $B_{0}$ is
aligned with the $y$-axis and $\mathrm{B}_{1}$ is in the $(x,z)$-plane,
such that:

\begin{alignat*}{1}
g(r) & =\sqrt{\left(\gamma_{\perp}\langle S_{x}\rangle\delta B_{1x}\right)^{2}+\left(\gamma_{\parallel}\langle S_{z}\rangle\delta B_{1z}\right)^{2}}
\end{alignat*}
For the nuclear-spin-free isotopes of $\text{Er}^{3+}$ and $\text{Yb}^{3+}$
(i.e: the $I=0$ transitions) the expectation values $\langle S_{x}\rangle$
and $\langle S_{z}\rangle$ can be calculated by diagonalising the
Zeeman Hamiltonian $H_{Z}=\hbar\boldsymbol{B_{0}}\cdot\boldsymbol{\gamma}\cdot\boldsymbol{S}$.
For the other Er electron-spin transitions, however, it's necessary
to add a hyperfine interaction term to account for the coupling of
the electronic-spin $S$ to the nuclear-spin $I$:
\begin{alignat}{1}
H & =\hbar\boldsymbol{B_{0}}\cdot\boldsymbol{\gamma}\cdot\boldsymbol{S}+\boldsymbol{S}\cdot\boldsymbol{A}\cdot\boldsymbol{I}\label{eq:Spin-Hamiltonian}
\end{alignat}
The hyperfine coupling matrix $A$ can be found in ref. \cite{Antipin1968}
and the resulting expectation values are tabulated below for the transitions
studied here.

\begin{table}[H]
\begin{centering}
\begin{tabular}{|c|c|c|c|c|}
\hline 
Transition & $B_{0}$ (mT) & $\langle S_{x}\rangle$ & $\langle S_{z}\rangle$ & $\frac{\tilde{\text{\ensuremath{\gamma}}}\hbar}{\mu_{B}}$\tabularnewline
\hline 
\hline 
$^{167}\text{Er}$ $|+\nicefrac{1}{2}\rangle$ & 34 & $-0.413i$ & 0.498 & 4.75\tabularnewline
\hline 
$\text{Er}^{3+}$ $I=0$ & 37 & $-i/2$ & $1/2$ & 5.71\tabularnewline
\hline 
$^{167}\text{Er}$ $|+\nicefrac{3}{2}\rangle$ & 43.5 & $-0.433i$ & $0.498$ & 4.98\tabularnewline
\hline 
$\text{Yb}^{3+}$ $I=0$ & 80 & $-i/2$ & $1/2$ & 2.75\tabularnewline
\hline 
\end{tabular}\caption{\label{tab:Sx-Sy-values}Expectation values $\langle S_{x}\rangle$
and $\langle S_{z}\rangle$ and `averaged' g-factors $\tilde{\text{g}}=\frac{\tilde{\text{\ensuremath{\gamma}}}\hbar}{\mu_{B}}$
for the ESR transitions relevant for determining the doping concentrations
of $\text{Er}^{3+}$ and $\text{Yb}^{3+}$. Transitions are listed
in order of increasing magnetic field $B_{0}$ in the ESR spectrum.}
\par\end{centering}
\end{table}

\subsection*{Subsection IV.2: Numerical estimate of the ensemble-coupling}

For an ensemble of spins with homogeneous doping concentration $\rho$
in a volume $V$, the ensemble-coupling $g_{ens}$ is enhanced by
a square-root factor, compared to the coupling of individual spins
\cite{Kubo2010}:

\begin{alignat*}{1}
g_{\text{ens}}= & \frac{\hbar}{\mu_{B}}\sqrt{\rho\int_{V}g(r)^{2}dr}
\end{alignat*}

This expression can be expressed as a function of several experimental
parameters. Firstly we can restrict the integral length along the
$y$-axis (the $b$-axis of the crystal) to the length $L$ of the
inductor shown in Figure \ref{fig:angles} (b). Moreover, we can exploit
the invariance of the magnetic field $\mathrm{B}_{1}$ along the $y$-axis
to further simplify the integral to just two spatially varying dimensions,
specifically $x$ and $z$:

\begin{alignat*}{1}
\int_{V}g(r)^{2}dr & =L\iint_{x<0}[\left(\gamma_{\perp}\langle S_{x}\rangle\delta B_{1x}\right)^{2}+\left(\gamma_{\parallel}\langle S_{z}\rangle\delta B_{1z}\right)^{2}]dxdz
\end{alignat*}

The domain can be further restricted to $x<0$ because the spins are
only located below the resonator and split the integral into two terms:
\begin{alignat*}{1}
\iint_{x<0}[\left(\gamma_{\perp}\langle S_{x}\rangle\delta B_{1x}\right)^{2}+\left(\gamma_{\parallel}\langle S_{z}\rangle\delta B_{1z}\right)^{2}]dxdz & =\frac{\iint_{x<0}[\left(\gamma_{\perp}\langle S_{x}\rangle\delta B_{1x}\right)^{2}+\left(\gamma_{\parallel}\langle S_{z}\rangle\delta B_{1z}\right)^{2}]dxdz}{\iint_{x<0}[\delta B_{1x}^{2}+\delta B_{1z}^{2}]dxdz}\iint_{x<0}[\delta B_{1x}^{2}+\delta B_{1z}^{2}]dxdz
\end{alignat*}

Here the first term is interpreted as an `averaged' gyromagnetic ratio,
which we label $\tilde{\text{\ensuremath{\gamma}}}$:
\begin{alignat*}{1}
\tilde{\text{\ensuremath{\gamma}}}= & 2\sqrt{\frac{\iint_{x<0}[\left(\gamma_{\perp}\langle S_{x}\rangle\delta B_{1x}\right)^{2}+\left(\gamma_{\parallel}\langle S_{z}\rangle\delta B_{1z}\right)^{2}]dxdz}{\iint_{x<0}[\delta B_{x}^{2}+\delta B_{z}^{2}]dxdz}}
\end{alignat*}

and the second term can be calculated by considering the energy of
an $n$-photon Fock state:

\begin{alignat*}{1}
E_{n} & =\hbar\omega_{r}(n+\frac{1}{2})\\
 & =\frac{1}{\mu_{0}}\int\langle n|\boldsymbol{\hat{B}_{1}}(\mathbf{r})^{2}|n\rangle d\mathbf{r}\\
 & =\frac{1}{\mu_{0}}\int\langle n|\boldsymbol{\delta B}(\mathbf{r})^{2}(\hat{a}+\hat{a}^{\dagger})^{2}|n\rangle d\mathbf{r}\\
 & =\frac{2n+1}{\mu_{0}}\int|\boldsymbol{\delta B}(\mathbf{r})|^{2}d\mathbf{r}
\end{alignat*}

which implies that:

\begin{alignat*}{1}
\int|\boldsymbol{\delta B}(\mathbf{r})|^{2}d\mathbf{r} & =\frac{\mu_{0}\hbar\omega_{r}}{2}
\end{alignat*}

Therefore second term becomes:
\begin{alignat*}{1}
\iint_{x<0}[\delta B_{x}^{2}+\delta B_{z}^{2}]dxdz & =\frac{1}{2L}\int|\boldsymbol{\delta B}(\mathbf{r})|^{2}d\mathbf{r}=\frac{\mu_{0}\hbar\omega_{r}}{4L}
\end{alignat*}

and finally we arrive at a simple numeric expression for the (temperature-independent)
ensemble coupling:

\begin{alignat}{1}
g_{\text{ens}} & =\frac{\tilde{\text{\ensuremath{\gamma}}}}{4}\sqrt{\rho\mu_{0}\hbar\omega_{r}}\label{eq:g_ens}
\end{alignat}

This equation, which has been previously determined in Ref \cite{Kubo2010},
links the ensemble coupling constant to the erbium concentration $\rho$.
The `averaged' spin-resonator coupling term$\tilde{\text{g}}$ is
computed using a COMSOL simulation of the magnetic field $\mathrm{B}_{1}$
over a $400\times200\,\mu m^{2}$ area representing $x\times z$ and
assuming a 5 $\mu m$ wide wire. This simulation includes the expectation
values of the spin-matrix elements presented in the third and fourth
columns of Table \ref{tab:Sx-Sy-values} and an example of such a
simulation is presented in Figure \ref{fig:angles} (c) for the Er
$I=0$ transition. The calculated values of $\tilde{\text{\ensuremath{\gamma}}}$
are presented in the fifth column of Table \ref{tab:Sx-Sy-values}.

\subsection*{\label{subsec:Sub-section-3:-Fitting}Subsection IV.3: Fitting the
temperature dependence of $\boldsymbol{g_{ens}}$}

Equation \ref{eq:g_ens} assumes complete polarisation of the electron-spins
within the ensemble. In reality the spin-polarisation $P$ is temperature
dependent, and Eq. \ref{eq:g_ens} should be adjusted to reflect this:

\begin{alignat}{1}
g_{\text{ens}}(T) & =\frac{\tilde{\text{\ensuremath{\gamma}}}}{4}\sqrt{P(T)\rho\mu_{0}\hbar\omega_{0}}\label{eq:g_ens-1}
\end{alignat}

\ref{subsec:Sub-section-3:-Fitting}

Assuming a Boltzmann thermal distribution, the thermal polarisation
of the nuclear-spin-free Er and Yb isotopes (i.e: the $I=0$ transitions)
is described by a hyperbolic tangent function: 

\begin{alignat*}{1}
P(T,I=0) & =\tanh\left(\frac{\hbar\gamma B_{0}}{2kT_{B}}\right)\\
 & =\tanh\left(\frac{\hbar\omega_{r}}{2kT_{B}}\right)
\end{alignat*}

where the usual thermal limits apply:
\begin{alignat*}{1}
\underset{T\rightarrow0}{lim}P(T,I=0) & =1\\
\underset{T\rightarrow\infty}{lim}P(T,I=0) & =0
\end{alignat*}

For the electron-spin transitions between hyperfine levels of $^{167}\text{Er}$,
however, one must consider all 16 hyperfine states defined by their
spin-projection $|m_{S},m_{I}\rangle$. This is necessary because
the small applied magnetic field used in these experiments can allow
all $\left(2S+1\right)\left(2I+1\right)=16$ hyperfine levels to be
populated at high temperatures. For a given electron-spin transition
of a nuclear-spin projection $|m_{I}\rangle$ the thermal polarisation
then follows the general Boltzmann form:

\begin{alignat}{1}
P(T,m_{I}) & =\frac{\exp[-E_{|+1/2,mI\rangle}/(kT)]-\exp[-E_{|-1/2,mI\rangle}/(kT)]}{\underset{\left(2S+1\right)\left(2I+1\right)}{\sum}\exp[-E_{|mS,mI\rangle}/(kT)]}\label{eq:themal pol}
\end{alignat}
where the energies $E_{|m_{S},m_{I}\rangle}$ is the energy of hyperfine
level $|m_{S},m_{I}\rangle$ at the magnetic field $B_{0}$ for which
$g_{ens}$ is measured. The fit to the $g_{ens}$ data presented in
Fig. 3 of the main text relies on additional knowledge of the relative
concentration of the$^{167}\text{Er}$ and $I=0$ Er isotopes, $\rho_{Er-167}$
and $\rho_{Er,I=0}$ respectively. In this way all three transitions
are fitted simultaneously with a single free parameter $\rho=\rho_{Er-167}+\rho_{Er,I=0}$
where:
\begin{alignat*}{1}
\rho & =\rho_{Er-167}+\rho_{Er,I=0}\\
 & =4.37\rho_{Er-167}\\
 & =1.3\rho_{Er,I=0}
\end{alignat*}

This yields an erbium concentration $\left[\text{Er}^{3+}\right]=2.26\times10^{17}\,\text{cm}^{-3}$
or equivalently 18 ppm relative doping. This is very similar to the
Er doping concentration determined by mass-spectrometry, presented
in Table I of the main text. 

\subsection*{Sub-section 4: Determining the $\boldsymbol{{\rm Yb}^{^{3+}}}$ concentration}

With an estimated erbium concentration of 18 ppm, it is possible to
further estimate the concentration of ytterbium by considering the
relative values of $\tilde{\gamma}$ and $g_{ens}$ of the nuclear-spin-free
($I=0$) ESR transitions for the two spin species:
\begin{alignat*}{1}
\frac{g_{ens}(\text{Er})}{g_{ens}(\text{Yb})} & =\frac{\tilde{\gamma}(\text{Er})}{\tilde{\gamma}(\text{Yb})}\frac{\sqrt{0.77\left[\text{Er}^{3+}\right]}}{\sqrt{0.7\left[\text{Yb}^{3+}\right]}}\\
\Rightarrow\left[\text{Yb}^{3+}\right] & =1.1\left[\text{Er}^{3+}\right]\left(\frac{g_{ens}(\text{Yb})\tilde{\gamma}(\text{Er})}{g_{ens}(\text{Er})\tilde{\gamma}(\text{Yb})}\right)^{2}\\
\text{substituting the known values gives}\\
\left[\text{Yb}^{3+}\right] & =1.29\times10^{17}\,\text{cm}^{-3}
\end{alignat*}

or equivalently, 10 ppm relative doping of $\text{Yb}^{3+}$.

\section*{Section V : Fitting the Three Pulse Echo data}

An accurate analysis of spectral diffusion in the millikelvin temperature
range requires that at least the Er and Yb $I=0$ sub-species be taken
into account in Eq. (2) of the main text. Both are required because
they exist in similar ppm concentrations (22.9 and 14.1 ppm, respectively)
while demonstrating very different gyromagnetic ratios and therefore
neither population is expected to dominate the spectral diffusion
processes across the entire millikelvin temperature range. This assumption
yields four spectral diffusion parameters : $\Gamma_{\text{SD}}^{Er},\,R^{Er},\,\Gamma_{\text{SD}}^{Yb}\,\&\,R^{Yb}$,
however, the physical similarity between these rare earth impurities
implies that the number of free parameters required to fit Eq. (2)
is no greater than for a single spin-species; namely two. In particular,
this fitting proceeds at each measured spin-bath temperature $T_{B}$
by assuming a relationship between the spectral diffusion linewidths
$\Gamma_{SD}$ governed by Eq. (3) of the main text: 
\begin{alignat*}{1}
\frac{\Gamma_{\text{SD}}^{Yb}}{\Gamma_{\text{SD}}^{Er}} & =\frac{\Gamma_{\text{max}}^{Yb}\text{sech}^{2}\left(g_{\text{Yb}}\mu_{B}B_{0}\left(2kT_{B}\right)^{-1}\right)}{\Gamma_{\text{max}}^{Er}\text{sech}^{2}\left(g_{\text{Er}}\mu_{B}B_{0}\left(2kT_{B}\right)^{-1}\right)}
\end{alignat*}

Where $\Gamma_{\text{max}}^{S}$ represents the maximum spectral-diffusion
linewidth, for which we assume an analytic form derived by Maryasov
et. al. \cite{Maryasov1982}:
\begin{alignat}{1}
\Gamma_{\text{max}}^{S} & =\frac{\mu_{B}^{2}\mu_{0}g_{\text{Er}}g_{\text{S}}n_{\text{S}}}{9\sqrt{3}\hbar}\label{eq:gamma_max}
\end{alignat}

Noting that the first $g$-factor is fixed for both species and determined
by the densest and most perturbative paramagnetic species in the matrix
(the erbium species), the spectral-diffusion contribution of the ytterbium
spins can be re-parametrised at each temperature $T_{B}$ as a function
of the erbium contribution $\Gamma_{\text{SD}}^{Er}$ and known impurity
concentrations $n_{\text{S}}$:
\begin{alignat}{1}
\Gamma_{\text{SD}}^{Yb} & =\Gamma_{\text{SD}}^{Er}\frac{g_{\text{Yb}}n_{\text{Yb}}\text{sech}^{2}\left(g_{\text{Yb}}\mu_{B}B_{0}\left(2kT_{B}\right)^{-1}\right)}{g_{\text{Er}}n_{\text{Er}}\text{sech}^{2}\left(g_{\text{Er}}\mu_{B}B_{0}\left(2kT_{B}\right)^{-1}\right)}\label{eq:gamma_Yb}
\end{alignat}

A similar treatment using Eq. (4) of the main text (and neglecting
the spin-lattice contribution) gives the following relationship for
the spin-flip rate $R$ :

\begin{alignat}{1}
R^{Yb} & =R^{Er}\frac{\Gamma_{\text{Er}}}{\Gamma_{\text{Yb}}}\frac{g_{\text{Yb}}^{4}n_{\text{Yb}}^{2}\text{sech}^{2}\left(g_{\text{Yb}}\mu_{B}B_{0}\left(2kT_{B}\right)^{-1}\right)}{g_{\text{Er}}^{4}n_{\text{Er}}^{2}\text{sech}^{2}\left(g_{\text{Er}}\mu_{B}B_{0}\left(2kT_{B}\right)^{-1}\right)}\label{eq:R_Yb}
\end{alignat}

where $\Gamma_{\text{Er}}=36.5$MHz \& $\Gamma_{\text{Yb}}=5.6$MHz
represent the inhomogeneous Er $I=0$ and Yb $I=0$ transition linewidths
respectively, measured at 100mK\footnote{See Figure 3 of the main text for examples the inhomogeneous erbium
linewidth measured at various temperatures.}. 

Using the relationships defined above it was possible to obtain unique
solutions to $\Gamma_{\text{SD}}^{Er}$ and $R^{Er}$(and by extension
$\Gamma_{\text{SD}}^{Yb}$ \& $R^{Yb}$) using Eq. (2) and fitting
to families of four (or more) three-pulse-echo decay curves at a fixed
temperature. Examples of this fitting for three different temperatures
are presented in Figure 4 of the main text.

\section*{Section VI : Decoherence rate model}

The model of decoherence rate (i.e homogeneous linewidth) $\Gamma_{h}$
presented in Fig. 5c of the main text is comprised of three contributions.
In particular, these are the instantaneous-diffusion decoherence rate
$\Gamma_{0}$, the electron-spin spectral diffusion decoherence rate
$\Gamma(SD)$ and the nuclear-spin spectral diffusion decoherence
rate $\Gamma(NSD)$. To combine these contributions we utilise the
$T_{2}$ formulation presented in Eq. 6 of the main text, together
with the decoherence-rate formalism $\Gamma_{h}=1/\pi T_{2}$,

\begin{alignat*}{1}
\Gamma_{h} & =\frac{2\Gamma(SD)^{2}+2\Gamma(NSD)^{2}}{-\Gamma_{0}+\sqrt{\Gamma_{0}^{2}+4\Gamma(SD)^{2}+4\Gamma(NSD)^{2}}}.
\end{alignat*}

The contribution $\Gamma_{0}$ is based on the analytic expression
presented in Chapter 3 of the PhD Thesis of M. Le Dantec \cite{LeDantec},
following the approach used by Tyryshkin \textit{et. al.} \cite{Tyryshkin2011}:

\begin{alignat*}{1}
\Gamma_{0} & =\frac{9\sqrt{3}\hbar}{\mu_{0}\left(g_{eff}\mu_{B}\right)^{2}}\frac{\Gamma_{|+\nicefrac{3}{2}\rangle}}{\Delta\omega}c_{|+\nicefrac{3}{2}\rangle}\sin^{-2}\left(\frac{\theta_{2}}{2}\right)
\end{alignat*}

Here $g_{eff}\approx g_{\perp}=8.38$ is the effective g-factor in
the direction of the applied field magnetic field. The ratio $\Gamma_{|+\nicefrac{3}{2}\rangle}/\Delta\omega$
represents the ratio of the inhomogeneous ESR transition linewidth
$\Gamma_{|+\nicefrac{3}{2}\rangle}\approx22$ MHz to pulse-exitation
bandwidth $\Delta\omega=700$ kHz. Furthermore, the spin-density $c_{|+\nicefrac{3}{2}\rangle}=P(T,|+\nicefrac{3}{2}\rangle)\rho_{167}$
represents the temperature-dependent density of the $m_{I}=|+\nicefrac{3}{2}\rangle$
hyperfine sub-level. It comprises the product of the spin-polarisation
$P(T,|+\nicefrac{3}{2}\rangle)$ given in Eq. \ref{eq:themal pol}
and the $^{167}\text{Er}$ defect density $\rho_{167}=5.2\cdot10^{16}\,\text{cm}^{-3}$,
determined from Table I of the main text. Finally, the angle $\theta_{2}=1.9$
rads is the average rabi-angle of the $\pi$ pulse used during echo
measurements: a value determined numerically for the inhomogeneous
$B_{1}$ field generated by the microwave resonator \cite{Ranjan2020a}.
Generally speaking, the decoherence contribution $\Gamma_{0}$ increases
with temperature and plateaus to a value of $650$ Hz at approximately
700mK, once all 16 hyperfine states of $^{167}\text{Er}$ are equally
populated. 

For the electron-spin contribution $\Gamma(SD)$ we consider only
the Er $I=0$ and Yb $I=0$ spin baths, and infer a temperature-dependent
rate:
\begin{alignat*}{1}
\Gamma(SD) & =\sqrt{\frac{1}{4\pi}\left(\Gamma_{\text{SD}}^{Er}R^{Er}+\Gamma_{\text{SD}}^{Yb}R^{Yb}\right)}
\end{alignat*}
for this contribution we further assume that the spectral-diffusion
linewidth $\Gamma_{\mathrm{SD}}^{S}$ and rate $R^{S}$ has a temperature
dependence following Eq. (3) and Eq. (4) of the main text: 

\begin{alignat*}{1}
\Gamma_{\text{SD}}^{S}R^{S}(T_{B}) & =\Gamma_{max}^{S}R_{max}^{S}\text{sech}^{4}\left(\frac{\hbar\gamma_{S}B_{0}}{2kT_{B}}\right)
\end{alignat*}

Here we determine the values of $\Gamma_{max}^{S}$ and $R_{max}^{S}$
using solely the analysis of the 3PE measurements. In particular,
we take the values presented by the dashed-line fits of Fig. 5a and
b of the main text. This yields $\Gamma_{max}^{Er}=400$ kHz, $R_{max}^{Er}=1.4\,\text{ms}^{-1}$
and $\Gamma_{max}^{Yb}=150$ kHz, $R_{max}^{Yb}=0.14\,\text{ms}^{-1}$
for Er for Yb contributions, respectively. 

Finally, we estimate the nuclear-spin contribution from the cluster-correlated-expansion
(CCE) simulation of the $^{183}\mathrm{W}$ nuclear-spin bath presented
in Fig. 2a of Ref. \cite{Dantec2021}:
\begin{alignat*}{1}
\Gamma(NSD) & =\frac{1}{\pi\cdot27\cdot10^{-3}\,s}\\
 & =12\,\text{Hz}
\end{alignat*}
 Although this CCE simulation was undertaken for the Er $I=0$ transition,
rather than the $m_{I}=|+\nicefrac{3}{2}\rangle$ transition, it is
still a valid estimate given the similarity of the transition-dipole
moments presented in Table \ref{tab:Sx-Sy-values} above. Note also
that this contribution is temperature independent because the $^{183}\mathrm{W}$
nuclear-spin bath remains unpolarised at millikelvin temperatures.

\bibliographystyle{plain}
\bibliography{C:/Mendeley_Bibtex/library}